\documentclass[11pt]{article}

\usepackage{authblk} 
\usepackage{graphicx}
\usepackage{epsfig}
\usepackage{dcolumn}
\usepackage{bm}
\usepackage{amsmath}
\usepackage{amssymb}

\newcommand{\beq}{\begin{equation}} \newcommand{\eeq}{\end{equation}}
\newcommand{\bea}{\begin{eqnarray}} \newcommand{\eea}{\end{eqnarray}}

  \newcommand
{\Romannumeral}[1]{\uppercase\expandafter{\romannumeral#1}}

\newcommand{\be}{\begin{enumerate}} \newcommand{\ee}{\end{enumerate}}
\newcommand{\bi}{\begin{itemize}} \newcommand{\ei}{\end{itemize}}
\newcommand{\ba}{\begin{array}} \newcommand{\ea}{\end{array}}
\newcommand{\bc}{\begin{center}} \newcommand{\ec}{\end{center}}
\newcommand{\bt}{\begin{tabular}} \newcommand{\et}{\end{tabular}}

\def\lsim{\mathrel{\rlap{\lower4pt\hbox{\hskip1pt$\sim$}}
    \raise1pt\hbox{$<$}}}           
\def\gsim{\mathrel{\rlap{\lower4pt\hbox{\hskip1pt$\sim$}}
    \raise1pt\hbox{$>$}}}           

\newcommand{\half}{\textstyle {1\over2} \displaystyle}    
\newcommand{\quarter}{\textstyle {1\over4} \displaystyle} 
\newcommand{\twoth}{\textstyle {2\over3} \displaystyle}   
\newcommand{\thrha}{\textstyle {3\over2} \displaystyle}   
\newcommand{\fivha}{\textstyle {5\over2} \displaystyle}   
\newcommand{\Dslash}{{\hbox{D}\kern-0.6em\raise0.15ex\hbox{/}}} 

\renewcommand{\et}{\eta}

\begin{document}

\setlength{\oddsidemargin}{0cm} \setlength{\baselineskip}{7mm}

\begin{normalsize}\begin{flushright}

July 2012 \\

\end{flushright}\end{normalsize}

\begin{center}
  
\vspace{5pt}

{\Large \bf Wheeler-DeWitt Equation in 2 + 1 Dimensions }

\vspace{30pt}

{\sl Herbert W. Hamber}$^{}$\footnote{e-mail address : HHamber@uci.edu},
{\sl Reiko Toriumi}$^{}$\footnote{e-mail address : RToriumi@uci.edu} \\
Department of Physics and Astronomy, \\
University of California, \\
Irvine, California 92697-4575, USA \\

\vspace{5pt}

and

\vspace{5pt}

{\sl Ruth M. Williams}
$^{}$\footnote{e-mail address : R.M.Williams@damtp.cam.ac.uk} \\
Department of Applied Mathematics and Theoretical Physics, \\
University of Cambridge,\\
Wilberforce Road, Cambridge CB3 0WA, United Kingdom, \\

and 

Girton College, University of Cambridge, Cambridge CB3 0JG, United Kingdom. \\

\vspace{10pt}

\end{center}

\begin{center} {\bf ABSTRACT } \end{center}

\noindent

The infrared structure of quantum gravity is explored by solving 
a lattice version of the Wheeler-DeWitt equations.
In the present paper only the case of $2+1$ dimensions is considered.
The nature of the wave function solutions is such that a finite correlation
length emerges and naturally cuts off any infrared divergences.
Properties of the lattice vacuum are consistent with the existence of an 
ultraviolet fixed point in $G$ located at the origin, thus precluding the 
existence of a weak coupling perturbative phase.
The correlation length exponent is determined exactly and
found to be $\nu=6/11$. 
The results obtained so far lend support to the claim that
the Lorentzian and Euclidean formulations belong to the same 
field-theoretic universality class.



\vfill

\pagestyle{empty}

\newpage

\pagestyle{plain}

\section{Introduction}
\label{sec:intro}

\vskip 20pt

It is possible that the well-known ultraviolet divergences affecting the
perturbative treatment
of quantum gravity in four dimensions point to a fundamental vacuum
instability of the full theory.
If this is the case, then the correct identification of the true ground
state for gravitation necessarily requires the introduction of a
consistent nonperturbative cutoff.
To this day the only known way to do this 
reliably in quantum field theory is via the lattice formulation.
Nevertheless, previous work on lattice quantum gravity has dealt almost exclusively
with the Euclidean formulation in $d$ dimensions, treated via
the manifestly covariant Feynman path integral method.
Indeed the latter is very well suited for numerical integration, and many
analytical and numerical results have been obtained over the years 
within this framework.
However the issue of their relationship to the Lorentzian theory
has remained largely open, at least from the point of view of a rigorous
treatment.
The main supporting arguments for the Euclidean approach come from
the fact that the above equivalence holds true for other field
theories (no exceptions are known), and from the fact that 
in gravity itself it is rigorously true to all orders in the weak field
expansion.

In this paper we will focus on the Hamiltonian approach to gravity,
which assumes from the beginning a metric with Lorentzian signature.
In order to obtain useful insights regarding the non-perturbative
ground state, a Hamiltonian lattice formulation was introduced based on the
Wheeler-DeWitt equation, where the quantum gravity Hamiltonian is 
written down in the position-space representation.
In a previous paper \cite{hw11} a general discrete Wheeler-DeWitt 
equation was given for pure gravity, based on the simplicial lattice 
formulation originally developed by Regge and Wheeler.
On the lattice the infinite-dimensional manifold of continuum
geometries is replaced by a finite manifold of piecewise
linear spaces, with solutions to the lattice equations then providing 
a suitable approximation to the continuum gravitational wave functional.
The lattice equations were found to be explicit enough to allow the
development of potentially useful practical solutions.
As a result, a number of sample quantum gravity calculations 
were carried out in $2+1$ and $3+1$ dimensions.
These were based mainly on the strong coupling expansion and on the
Rayleigh-Ritz variational method, the latter implemented using 
a set of correlated product (Slater-Jastrow) wave functions.  

Here, we extend the work initiated in \cite{hw11} and
show how exact solutions to the lattice Wheeler-DeWitt
equations can be obtained in $ 2 + 1 $ dimensions
for arbitrary values of Newton's constant $G$.
The procedure we follow is to solve the lattice equations exactly for
several finite regular triangulations of the sphere and then 
extend the result to an arbitrarily large number of triangles.
One finds that for large enough areas the exact lattice wave functional depends
on geometric quantities only, such as the total area and the total
integrated curvature (which in $ 2 + 1$ dimensions is just proportional
to the Euler characteristic).
The regularity condition on the solutions of the wave equation at
small areas is shown to play an essential role in constraining the form of the
wave functional, which we eventually find to be expressible in closed form 
as a confluent hypergeometric function of the first kind.
Later it will be shown that the resulting wave function allows an exact evaluation
of a number of useful (and manifestly diffeomorphism-invariant)
averages, such as the average area of the manifold and its
fluctuation.

From these results a number of suggestive physical results can be obtained,
the first one of which is that the correlation length in units of the 
lattice spacing is found to be finite for all $G>0$, and diverges at $G=0$.
Such a result can be viewed as consistent with the existence of an 
ultraviolet fixed point (or a phase transition in statistical field theory language)
in $G$ located at the origin, thus entirely precluding the existence of 
a weak coupling phase for gravity in $2+1$ dimensions.
Simple renormalization group arguments would then suggest that
gravitational screening is not physically possible in $2+1$
dimensions, and that gravitational antiscreening is the only 
physically realized option in this model.
A second result that follows from our analysis is an exact
determination of the critical correlation length exponent for 
gravity in $2+1$ dimensions, which is found to be $\nu=6/11$. 
It is known that the latter determines, through standard 
renormalization group arguments, the scale dependence of the 
gravitational coupling in the vicinity of the ultraviolet fixed point.

A short outline of the paper is as follows.
In Sec. 2, as a general background to the rest of the paper, 
we briefly describe the formalism of classical canonical gravity, 
as originally formulated by Arnowitt, Deser and Misner.
The continuum Wheeler-DeWitt equation and its invariance
properties are introduced as well at this stage.
In Sec. 3 we introduce the lattice Wheeler-DeWitt equation
derived in a previous paper \cite{hw11},
and later Sec. 4 makes more explicit various quantities 
appearing in it.
This last section also discusses briefly the role of continuous lattice
diffeomorphism invariance in the Regge framework
as it applies to the present case of $2+1$-dimensional gravity.
Sec. 5 focuses on the scaling properties of the lattice equations
and various sensible choices for the lattice coupling constants,
with the aim of giving eventually a more transparent form to the 
wave function results.
Sec. 6 gives a detailed outline of the general method 
of solution for the lattice equations and then gives the 
explicit solution for a number of regular triangulations
of the sphere.
Later, a general form of the wave function is given that covers all
the previous discrete cases and allows a subsequent study of the 
infinite volume limit.
Sec. 7 focuses on one of the simplest diffeomorphism-invariant
averages that can be computed from the wave function, namely
the average total area.
A brief discussion follows on how the latter quantity relates to 
the corresponding averages computed in the Euclidean theory.
Sec. 8 extends the calculation to the area fluctuation and shows
how the critical exponents (anomalous dimensions) of the $2+1$-gravity
theory can be obtained from the exact wave function solution, 
using some rather straightforward scaling arguments.
Sec. 9 discusses some simple physical implications that can be inferred
from the values of the exact exponents and the fact that quantum 
gravity in $2+1$ dimensions does not seemingly possess, in either the
Euclidean or Lorentzian formulation, a weak coupling phase.
Sec. 10 contains a summary of the results obtained so far.

\vskip 40pt

\section{Continuum Wheeler-DeWitt Equation}
\label{sec:wdweq}

Since this paper involves the canonical quantization of gravity 
we begin here with a very brief summary of the classical canonical formalism
\cite{dir58}
as derived by Arnowitt, Deser and Misner 
\cite{adm62}.
While many of the results presented in this section are rather well known, it will be
useful, in view of later applications, to recall the main results and formulas and
provide suitable references for expressions used later in the paper.

The first step in developing a canonical formulation for gravity is to 
introduce a time-slicing of space-time, by introducing a sequence of 
spacelike hypersurfaces labeled by a continuous time coordinate $t$.
The invariant distance is then written as
\beq
d s^2 \; \equiv \; - d \tau^2 \, = \, g_{\mu\nu} \, dx^\mu dx^\nu 
 \, = \, g_{ij} \, d x^i  \, d x^j  \, + \, 2 g_{ij} \, N^i dx^j dt 
\, - \, ( N^2 \, - \, g_{ij} \, N^i N^j ) dt^2 \; ,
\eeq
where 
$x^i$ $(i=1,2,3)$ 
are coordinates on a three-dimensional manifold and $\tau$ 
is the proper time, in units with $c=1$.

Indices are raised and lowered with 
$g_{ij} ( {\bf x} ) $ $(i,j=1,2,3)$, 
which denotes the three-metric on the given spacelike hypersurface, and 
$N( {\bf x} )$ 
and 
$N^i ( {\bf x} )$ 
are the lapse and shift functions, respectively.
It is customary to mark four-dimensional quantities by the prefix 
${}^4$
so that all unmarked quantities will refer to three dimensions 
(and are occasionally marked explicitly by a ${}^3$).
In terms of the original four-dimensional metric 
${}^4 g_{\mu\nu}$ 
one has
\beq
\left ( 
\begin{matrix}
{}^4 g_{00} & {}^4 g_{0j} \cr
{}^4 g_{i0} & {}^4 g_{ij}  
\end{matrix}  \right )
\; = \; 
\left ( 
\begin{matrix}
N_k N^k - N^2 & N_j  \cr
N_i  & g_{ij} 
\end{matrix}  \right )  \;\; ,
\eeq
which then gives for the spatial metric and the lapse and shift functions
\beq
g_{ij} \; = \; {}^4 g_{ij} \;\;\;\;\;\;
N \; = \;  \left ( - {}^4 g^{00} \right )^{-1/2} \;\;\;\;\;\;
N_i \; = \; {}^4 g_{0i} \; .
\eeq
For the volume element one has
\beq
\sqrt{ \, - \, {}^4 g } \; = \; N \, \sqrt{g} \; ,
\eeq
where the latter involves the determinant of the three-metric,
$g \equiv \det g_{ij}$.
As usual 
$g^{ij}$ 
denotes the inverse of the matrix 
$g_{ij}$.

A transition from the classical to the quantum description 
of gravity is obtained by promoting the metric
$g_{ij}$, the conjugate momenta $\pi^{ij}$, 
the Hamiltonian density $H$ and the momentum density $H_i$ 
to quantum operators, with 
${\hat g}_{ij}$ 
and 
${\hat \pi}^{ij}$ 
satisfying canonical commutation relations.
In particular, the classical constraints now select a physical vacuum state 
$ \vert \Psi \rangle $, 
such that in the source-free case
\beq
{\hat H} \, \vert \Psi \rangle \; = \; 0
\;\;\;\;\;\;
{\hat H}_i \, \vert \Psi \rangle \; = \; 0 \;\; 
\eeq
and in the presence of sources more generally
\beq
{\hat T} \, \vert \Psi \rangle \; = \; 0
\;\;\;\;\;\;
{\hat T}_i \, \vert \Psi \rangle \; = \; 0 \;\; ,
\label{eq:quant_const}
\eeq
where ${\hat T}$ and ${\hat T}_i$ now include matter
contributions that should be added to ${\hat H}$ and ${\hat H}_i$.
The momentum constraint involving $ {\hat H}_i $ or
more generally $ {\hat T}_i $, ensures that the state
functional does not change under a transformation
of coordinates $x^i$, so that $\Psi$ depends only on
the intrinsic geometry of the 3-space.
The Hamiltonian constraint is then the only remaining
condition that the state functional must satisfy.

As in ordinary nonrelativistic quantum mechanics, one can choose 
different representations for the canonically conjugate operators 
${\hat g}_{ij}$ and ${\hat \pi}^{ij}$.
In the functional position representation one sets
\beq
{\hat g}_{ij} ( {\bf x} ) \; \rightarrow \;   g_{ij} ( {\bf x} ) 
\;\;\;\;\;\;
{\hat \pi}^{ij} ( {\bf x} ) \; \rightarrow \;  
- i \hbar \cdot16 \pi G \cdot
{ \delta \over \delta g_{ij} ( {\bf x} )  } \; .
\eeq
In this picture quantum states become wave functionals of the 
three-metric 
$ g_{ij} ({\bf x} ) $,
\beq
\vert \Psi \rangle \; \rightarrow \;  \Psi \, [ g_{ij} ( {\bf x} ) ] \; .
\eeq
The two quantum-constraint equations in Eq.~(\ref{eq:quant_const})
then become the Wheeler-DeWitt equation \cite{dew64, dew67, whe68}
\beq
\left \{ -  \, 16 \pi G \cdot G_{ij,kl}  \, 
{ \delta^2 \over \delta g_{ij} \, \delta g_{kl}  } \; - \;
{ 1 \over 16 \pi G } \, \sqrt{g} \left ( \,  {}^3 \! R \, - \, 2 \lambda \, \right )
\, + \,  {\hat H}^\phi \right \} \; \Psi [ g_{ij} ( {\bf x} ) ] \; = \; 0 \; ,
\label{eq:wd_1}
\eeq
and the momentum constraint listed below.
Here $ G_{ij,kl} $ is the inverse of the DeWitt supermetric, given by
\beq
G_{ij,kl} \; = \;  \half \, g^{-1/2} \left ( 
g_{ik} g_{jl} + g_{il} g_{jk} + \alpha \, g_{ij} g_{kl} \right ) \;\; ,
\label{eq:dewitt-metric3d-inv}
\eeq
with parameter $\alpha = -1 $.
The three-dimensional version of the DeWitt supermetric itself,
$ G^{ij,kl} (x) $ is given by
\beq
G^{ij,kl} \; = \;  
\half \, \sqrt{g} \, \left ( g^{ik} g^{jl} + g^{il} g^{jk} 
+ \bar{\alpha} \, g^{ij} g^{kl} \right ) \;\; ,
\label{eq:dewitt-metric3d}
\eeq
with parameter $\alpha$ in Eq.~(\ref{eq:dewitt-metric3d-inv})
related to $\bar{\alpha}$ in Eq.~(\ref{eq:dewitt-metric3d}) by 
$\bar{\alpha} = - 2 \alpha / ( 2 + 3 \alpha ) $, 
so that
$\alpha = -1 $ 
gives 
$\bar{\alpha} = -2$ 
(note that this is dimension dependent). 
In the position representation the diffeomorphism (or momentum)
constraint reads
\beq
\left \{ 2 \, i  \, g_{ij}  \, \nabla_k \, { \delta \over \delta g_{jk}  } \, + \,
{\hat H}^\phi_i \right \} \; \Psi [ g_{ij} ( {\bf x} )  ] \; = \; 0 \; ,
\label{eq:wd_2}
\eeq
where $ {\hat H}^\phi$ and $ {\hat H}^\phi_i $ are possible matter
contributions.
In the following, we shall set both of these to zero, as we will focus here
almost exclusively on the pure gravitational case.

A number of basic issues need to be addressed before one can 
gain a full and consistent understanding of the dynamical content of the theory 
(see, for example, \cite{leu64,kuc76,kuc92,ish93,bar98} as a small set of
representative references).
These include possible problems of operator ordering, 
and the specification of a suitable Hilbert space, 
which entails at some point a choice for the inner product 
of wave functionals,  for example in the Schr\"odinger form
\beq
\langle \Psi \vert \Phi \rangle \; = \; 
\int  d \mu [g] \; \Psi^{*} [ g_{ij} ] \; \Phi [ g_{ij} ] \,
\label{eq:norm}
\eeq
where $ d \mu [g] $ is some appropriate measure over the three-metric $g$.
Note also that the continuum Wheeler-DeWitt equation contains,
in the kinetic term, products of functional differential operators 
which are evaluated at the same spatial point 
${\bf x}$.
One would expect that such terms could produce 
$\delta^{(3)} (0)$
-type singularities when acting on the wave functional, which 
would then have to be regularized in some way.
The lattice cutoff discussed below is one way to provide such an
explicit ultraviolet regularization.

A peculiar property of the Wheeler-DeWitt equation, 
which distinguishes it from the usual Schr\"odinger equation
$H \Psi = i \hbar \, \partial_t \Psi $, 
is the absence of an explicit time coordinate.
As a result, the rhs term of the Schr\"odinger equation is here entirely absent.
The reason is of course diffeomorphism invariance of the underlying theory, 
which expresses now the fundamental quantum equations in terms of fields 
$g_{ij}$ and not coordinates.

\vskip 40pt

\section{Lattice Hamiltonian for Quantum Gravity}
\label{sec:ham-grav-latt}

In constructing a discrete Hamiltonian for gravity, one has to decide 
first what degrees of freedom one should retain on the lattice.
One possibility, which is the one we choose to pursue here, 
is to use the more economical (and geometric) Regge-Wheeler lattice discretization for gravity 
\cite{reg61,whe64},
with edge lengths suitably defined on a random lattice as the primary dynamical variables.
Even in this specific case several avenues for discretization are possible.
One could discretize the theory from the very beginning, 
while it is still formulated in terms of an action, and introduce 
for it a lapse and a shift function, extrinsic and intrinsic discrete curvatures etc.
Alternatively one could try to discretize the continuum Wheeler-DeWitt 
equation directly, a procedure that makes sense in the lattice formulation, 
as these equations are still given in terms of geometric objects, 
for which the Regge theory is very well suited.
It is the latter approach which we will proceed to outline here.

The starting point for the following discussion is therefore the 
Wheeler-DeWitt equation for pure gravity in the absence of matter, 
Eq.~(\ref{eq:wd_1}), 
\beq
\left \{ \, -  \, (16 \pi G )^2 \, G_{ij,kl}  ( {\bf x} )  \, 
{ \delta^2 \over \delta g_{ij} ( {\bf x} ) \, \delta g_{kl}  ( {\bf x} ) } \; - \;
\sqrt{g ( {\bf x} ) } \; \left ( \;  
{}^3 \! R ( {\bf x} ) \, - \, 2 \lambda \, \right ) \, \right \} \; 
\Psi [ g_{ij} ( {\bf x} ) ] \; = \; 0
\label{eq:wd_1a}
\eeq
and the diffeomorphism constraint of 
Eq.~(\ref{eq:wd_2}),
\beq
\left \{ \, 2 \, i  \, g_{ij}  ( {\bf x} ) \, \nabla_k ( {\bf x} ) \,
 { \delta \over \delta g_{jk}  ( {\bf x} ) }  \, \right \} \; 
\Psi [ g_{ij} ( {\bf x} )  ] 
\; = \; 0 \; .
\label{eq:wd_2a}
\eeq
Note that these equations express a constraint on the state 
$ \vert \Psi \rangle$
at {\it every} 
${\bf x}$, 
each of the form 
$ {\hat H} ( {\bf x} ) \, \vert \Psi \rangle = 0$ 
and
$ {\hat H}_i \, ( {\bf x} ) \vert \Psi \rangle = 0 $.

On a simplicial lattice 
\cite{rowi81,che82,itz83,hw84,har85} 
(see for example
\cite{hbook}, 
and references therein, for a more complete discussion of the lattice formulation for gravity)
one knows that deformations of the squared edge lengths are linearly 
related to deformations of the induced metric.
In a given simplex $\sigma$, take coordinates based at a vertex $0$, 
with axes along the edges from $0$.
The other vertices are each at unit coordinate distance from $0$ 
(see Figs. 1, 2 and 3 as an example of this labeling for a triangle). 
In terms of these coordinates, the metric within the simplex is given by
\beq
g_{ij} (\sigma) \; = \; \half \, 
\left ( l_{0i}^2 + l_{0j}^2 - l_{ij}^2 \right ) \;\; .
\label{eq:latmet-1}
\eeq
Note that in the following discussion only edges and 
volumes along the spatial direction are involved.
It follows that one can introduce in a natural way a 
lattice analog of the DeWitt supermetric of 
Eq.~(\ref{eq:dewitt-metric3d})
by adhering to the following procedure \cite{lun74,har97}.
First one writes for the supermetric in edge length space 
\beq
\Vert \, \delta l^2 \, \Vert^2 \; = \; \sum_{ij} \; G^{ij} (l^2)
\; \delta l^2_i \; \delta l^2_j \; \; ,
\eeq
with the quantity 
$G^{ij} (l^2)$ 
suitably defined on the space of squared edge lengths.
By a straightforward exercise of varying the squared volume of a given simplex 
$\sigma$ 
in 
$d$
dimensions
\beq
V^2 ( \sigma ) \; = \; {\textstyle \left ( { 1 \over d! } \right )^2 \displaystyle}
\det g_{ij}(l^2( \sigma )) 
\eeq
to quadratic order in the metric (on the rhs), or in the squared 
edge lengths belonging to that simplex (on the lhs), one is led to
the identification
\beq
G^{ij} (l^2) \; = \; - \; d! \; \sum_{\sigma} \;
{1 \over V (\sigma)} \; 
{ \partial^2 \; V^2 (\sigma) \over \partial l^2_i \; \partial l^2_j } \;\; .
\label{eq:inverselasup}
\eeq
It should be noted that in spite of the appearance of a sum over simplices 
$\sigma$, $G^{ij} (l^2) $ 
is local, since the sum over $\sigma$ only extends over those simplices
which contain either the $i$ or the $j$ edge.

At this point one is finally ready to write a lattice analog of 
the Wheeler-DeWitt equation for pure gravity, which reads 
\beq
\left \{ \, -  \, (16 \pi G)^2 \, G_{ij} ( l^2 ) \, 
{ \partial^2 \over \partial l^2_{i} \, \partial l^2_{j}  }
\; - \;
\sqrt{g (l^2) } \; \left [ \; 
{}^3 \! R (l^2) \, - \, 2 \lambda \; \right ] \; \right \} \; 
\Psi [ \, l^2 \, ] \; = \; 0 \;\; ,
\label{eq:wd_latt}
\eeq
with $G_{ij} (l^2)$ the inverse of the matrix $G^{ij} (l^2)$ given above.
The range of the summation over $i$ and $j$ 
and the appropriate expression for the scalar curvature, 
in this equation, are discussed below and made explicit in 
Eq.~(\ref{eq:wd_latt1}).

Equations (\ref{eq:wd_1}) or (\ref{eq:wd_latt})
express a constraint equation at each \lq\lq point" in space.
Here we will elaborate a bit more on this point.
On the lattice, points in space are replaced by a set of edge labels $i$, 
with a few  edges clustered around each vertex in a way that 
depends on the dimensionality and the local lattice coordination number.
To be more specific, the first term in Eq.~(\ref{eq:wd_latt})
contains derivatives with respect to edges $i$ and $j$ 
connected by a matrix element $ G_{ij} $ which is nonzero only if $i$ and $j$
are close to each other, essentially nearest neighbor.
One would therefore expect that the first term could be represented
by just a sum of edge contributions, all from within one
($d-1$)-simplex $\sigma$ (a tetrahedron in three dimensions).
The second  term containing ${}^3 \! R (l^2)$ in Eq.~(\ref{eq:wd_latt}) 
is also local in the edge lengths:
it only involves a handful of edge lengths, which enter 
into the definition of areas, volumes and angles around the point 
${\bf x}$, and follows from the fact that the local curvature at 
the original point ${\bf x}$ is completely determined by the values
of the edge lengths clustered around $i$ and $j$.
Apart from some geometric factors, it describes, through a deficit angle 
$\delta_h$, 
the parallel transport of a vector around an elementary dual lattice loop.
It should, therefore, be adequate to represent this second term 
by a sum over contributions over all ($d-3$)-dimensional hinges 
(edges in 3+1 dimensions) $h$ attached to the simplex $\sigma$,
giving, therefore, in three dimensions 
\beq
\left \{ \, -  \, (16 \pi G )^2 \sum_{ i,j \subset \sigma}
\, G_{ij} \, ( \sigma ) \, 
{ \partial^2 \over \partial l^2_{i} \, \partial l^2_{j}  }
\; - \; 2 \, n_{\sigma h} \; \sum_{ h \subset \sigma} \, l_h \, \delta_h 
\; + \; 2 \lambda  \; V_\sigma  \,
\right \} \; 
\Psi [ \, l^2 \, ] \; = \; 0 \;\; .
\label{eq:wd_latt1}
\eeq
Here $\delta_h$ is the deficit angle at the hinge $h$, $l_h$ the corresponding edge length, 
and $V_\sigma = \sqrt{ g(\sigma)} $ the volume of the simplex (tetrahedron
in three spatial dimensions)
labeled by $\sigma$.
$ G_{ij} \, ( \sigma ) $ is obtained either from Eq.~(\ref{eq:inverselasup})
or from the lattice transcription of Eq.~(\ref{eq:dewitt-metric3d-inv})
\beq
G_{ij,kl} (\sigma) \; = \;  \half \, g^{-1/2} (\sigma) \left [ 
g_{ik} (\sigma) g_{jl} (\sigma) + g_{il} (\sigma) g_{jk} (\sigma) - 
\, g_{ij} (\sigma) g_{kl} (\sigma) 
\right ] \; \; ,
\label{eq:dewitt-metric3d-inv-latt}
\eeq
with the induced metric $g_{ij} (\sigma) $ within a simplex $\sigma$ given in 
Eq.~(\ref{eq:latmet-1}).
The combinatorial factor $ n_{\sigma h} $ ensures the correct
normalization for the curvature term, since the latter has to give the lattice version of
$ \int \sqrt{g} \, {}^3 R = 2 \sum_h \delta_h l_h $  
(in three spatial dimensions) when summed over all simplices 
$\sigma$.
The inverse of $ n_{\sigma h} $ counts, therefore, the number of times the same hinge appears in 
various neighboring simplices and consequently depends on the 
specific choice of underlying lattice structure; 
for a flat lattice of equilateral triangles in two dimensions, 
$ n_{\sigma h} =1/6 $.\footnote{
Instead of the combinatorial factor $n_{\sigma h} $, one could insert a ratio of volumes 
$V_{\sigma h}/ V_h$ (where $V_h$ is the volume per hinge \cite{hw84} 
and $V_{\sigma h}$ is the amount of that volume in the simplex $\sigma$), 
but the above form is simpler.}
The lattice Wheeler-DeWitt equation given in Eq.~(\ref{eq:wd_latt1})
was the main result of a previous paper \cite{hw11}.


\vskip 40pt

\section{Explicit Setup for the Lattice Wheeler-DeWitt Equation}
\label{sec:calc}

In this section, we shall establish our notation and derive the 
relevant terms in the discrete Wheeler-DeWitt equation for a simplex.
From now on we shall focus almost exclusively on the case of 2+1
dimensions.
The basic simplex in this case is, of course, a triangle, 
with vertices and squared edge lengths labelled as in Fig. 1. 
We set 
$l_{01}^2=a, \, l_{12}^2=b, \, l_{02}^2=c$.
The components of the metric for coordinates based at vertex $0$, 
with axes along the $01$ and $02$ edges, are
\beq
g_{11} \; = \; a , \;\;\;\;\; g_{12} \; = \; \frac{1} {2} (a + c - b),
\;\;\;\; g_{22} \; = \; c .
\eeq
The area $A$ of the triangle is given by
\beq
A^2 \; = \; 
\frac {1} {16} \, 
[ \, 2 ( ab + bc + ca ) - a^2 - b^2 -  c^2 \, ] \; ,
\eeq
so the supermetric $G^{ij}$, according to Eq.~(\ref{eq:inverselasup}), is
\beq
G^{ij} \; = \; \frac {1} {4A} \;\;
\left (  \begin{matrix}
1 & -1 & -1 \cr
-1 & 1 & -1 \cr
-1 & -1 & 1
\end{matrix}  \right  ) ,
\eeq
Thus for the triangle we have
\beq
G_{ij} \; {\partial^2 \over \partial s_i \; \partial s_j} \; = \; - 4 A
\left ( {\partial^2 \over \partial a \; \partial b} \; + \; {\partial^2
\over \partial b \; \partial c} \; + \; {\partial^2 \over \partial c \;
\partial a} \right ),
\label{eq:laplace_2d}
\eeq
and the Wheeler-DeWitt equation is  
\beq
\left  \{  (16 \pi G)^2 \; 4 A \; \left ( {\partial^2 \over \partial a
\; \partial b} \; + \; {\partial^2 \over \partial b \; \partial c} \;
+ \; {\partial^2 \over \partial c \; \partial a} \right ) \; - \;
2 \; n_{\sigma h} \, \sum_h \delta_h \; + \; 2 \lambda A \right \} \Psi
[\, s \,] \; = \; 0,
\label{eq:wd_tria}
\eeq
where the sum is over the three vertices $h$ of the triangle. 

\begin{figure}
\begin{center}
\includegraphics[width=.6\textwidth]{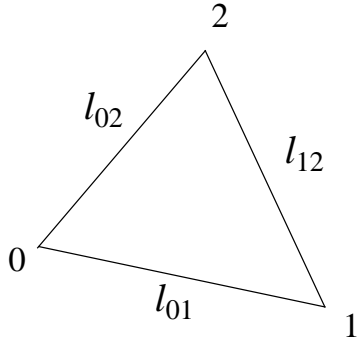}
\end{center}
\caption{\label{fig:triangle}A triangle with labels. }
\end{figure}

\begin{figure}
\begin{center}
\includegraphics[width=.5\textwidth]{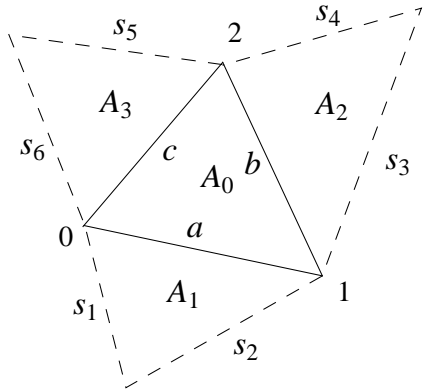}
\end{center}
\caption{\label{fig:neighbors} 
Neighbors of a given triangle.
The picture illustrates the fact that the Laplacian 
${\bf \Delta} (l^2)$ 
appearing in the kinetic term of the lattice Wheeler-DeWitt 
equation (here in 2+1 dimensions) contains edges 
$a,b,c$ 
that belong both to the triangle in question, 
as well as to several neighboring triangles (here three of them) 
with squared edges denoted sequentially by 
$s_1 =l_1^2 \dots s_6=l_6^2 $.}
\end{figure}

\begin{figure}
\begin{center}
\includegraphics[width=.6\textwidth]{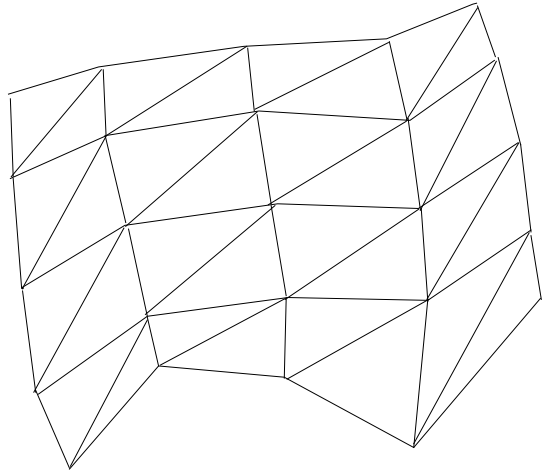}
\end{center}
\caption{\label{fig:lattice2d} A small section of a suitable dynamical
spatial lattice for quantum gravity in 2+1 dimensions.}
\end{figure}

In the following sections we will be concerned at some point with various discrete,
but generally regular, triangulations
of the two-sphere, such as the tetrahedron, the octahedron and the
icosahedron.
These were already studied in some detail in \cite{gauge,hw86}.
A key aspect of the Regge theory is the presence of a continuous, local
lattice diffeomorphism invariance, whose main aspects in regard
to their relevance for the $3+1$ formulation of gravity were
already addressed in some detail in \cite{hw11} in the context of
the lattice weak field expansion.
Here we will add some remarks about how this local invariance
manifests
itself in the $2+1$ formulation, and, in particular, for the discrete
triangulations of the sphere studied later on in this paper.
Of some relevance is the presence
of exact zero modes of the gravitational lattice action, reflecting
a local lattice diffeomorphism invariance, present already on
a finite lattice.
Since the Einstein action is a topological invariant in two dimensions,
the relevant action in this case has to be a curvature-squared action
supplemented by a cosmological constant term.
Specifically, part of the results in \cite{hw93,gauge} can be summarized 
as follows.
For a given lattice, one finds for the counting of zero modes
\bea
{\rm Tetrahedron} \; (N_0 = 4) : && {\rm 2 \; zero \; modes}
\nonumber \\
{\rm Octahedron} (N_0 = 6) : && {\rm 6 \; zero \; modes}
\nonumber \\
{\rm Icosahedron} (N_0 = 12 ) : && {\rm 18 \; zero \; modes} \; .
\nonumber \\
\eea
Thus if the number of zero modes for each regular triangulation of the sphere is
denoted by $N_{z.m.}$, then the results can be reexpressed as
\beq
N_{z.m.} \; = \; 2 \, N_0 \, - \, 6 \;\; ,
\eeq
which agrees with the expectation that, in the continuum limit,
$N_0 \rightarrow \infty$, $N_{z.m.}/N_0$ should approach the constant
value $d$ in $d$ space-time dimensions,
the expected number of local parameters for a diffeomorphism.
Similar estimates were obtained when looking at deformations 
of a flat lattice in various dimensions \cite{gauge}.
The case of near-flat space is obviously the simplest: by moving the
location of the vertices around in flat space, one can find a different
assignment of edge lengths that represents the same flat geometry.
This then leads to the $ d \cdot N_0$-parameter family of transformations for the
edge lengths in flat space.

In general, lattice diffeomorphisms correspond to local deformations
of the edge lengths about a vertex, which leave the local geometry
physically unchanged, the latter being described by the values of local
lattice operators corresponding to local volumes, curvatures, etc.
The lesson is that the correct count of continuum zero modes will, in general, only be
recovered asymptotically for large triangulations, where $N_0 $ is
significantly larger than the number of neighbors to a point in $d$
dimensions.
With these observations in mind, we can now turn to a discussion
of the solution method for the lattice Wheeler-DeWitt equation
in $2+1$ dimensions.

One item that needs to be discussed at this point is the proper
normalization of various terms (kinetic, cosmological and curvature)
appearing in the lattice equation of Eq.~(\ref{eq:wd_latt1}).
For the lattice gravity action in $d$ dimensions one has generally
the following correspondence
\beq
\int d^d x \, \sqrt{g}  \;\;\; \longleftrightarrow  \;\;\;  \sum_{\sigma} V_{\sigma}
\label{eq:reg_v}
\eeq
where $V_\sigma$ is the volume of a simplex; in two dimensions it is
simply the area of a triangle.
The curvature term involves deficit angles in the discrete case,
\beq
\half \,  \int  \, d^d x \, \sqrt{g} \, R  \;\;\;  \longleftrightarrow
 \;\;\;   \sum_{h} V_h \, \delta_h
\label{eq:reg_r}
\eeq
where $ \delta_h $ is the deficit angle at the hinge $h$, and $V_h$ the
associated ``volume of the hinge'' \cite{reg61}.
In four dimensions the latter is the area of a triangle (usually
denoted by $A_h$), whereas in three dimensions it is simply given by the
length $l_h$ of the edge labeled by $h$. 
In two dimensions $V_h = 1$.
In this work we will focus almost exclusively on the case of $2+1$ dimensions;
consequently the relevant formulas will be 
Eqs.~(\ref{eq:reg_v}) and (\ref{eq:reg_r}) for dimension $d=2$.

The continuum Wheeler-DeWitt equation is local, as can
be seen from Eq.~(\ref{eq:wd_1a}).
One can integrate the Wheeler-DeWitt operator over all space and obtain
\beq
\left \{ - \, \left( 16 \pi \, G \right)^2 \, 
\int d^2x \; {\bf \Delta} (g) + 2 \lambda \int d^2x \, \sqrt{g} 
- \int d^2x \, \sqrt{g} \, R \right \} \, \Psi = 0
\eeq
with the super-Laplacian on metrics defined as
\beq
{\bf \Delta } (g) \;  \equiv  \; G_{ij,kl}  ( {\bf x} )  \, 
{ \delta^2 \over \delta g_{ij} ( {\bf x} ) \, 
\delta g_{kl} ( {\bf x}  ) } \; .
\label{eq:lapl}
\eeq
In the discrete case one has one local Wheeler-DeWitt equation for
{\it each} triangle 
[see Eqs.~(\ref{eq:wd_latt}) and (\ref{eq:wd_latt1})],
which therefore takes the form
\beq
\left \{ \, - \, ( 16 \pi \, G)^2 \, {\bf \Delta }( l^2 ) 
- \, \kappa \, \sum_{ i \subset \Delta } \delta_i 
+ 2 \, \lambda \, A_{\Delta}  \,
\right \} \, \Psi = 0 \; ,
\label{eq:wd_2d}
\eeq
where ${\bf \Delta }( l^2 )$ is the lattice version of the
super-Laplacian, and we have set for convenience $ \kappa = 2 \, n_{\sigma \, h} $.
As we shall see below, for a lattice of fixed coordination number,
$\kappa$ is a constant and does not depend on the location
on the lattice.
In the above expression ${\bf \Delta} ( l^2 )$ is a discretized form 
of the covariant super-Laplacian, 
acting locally on the space of $s=l^2$ variables.
From Eqs.~(\ref{eq:laplace_2d}) and (\ref{eq:wd_2d}) one has explicitly
\beq
{\bf \Delta }( l^2 ) \;  = \; - \, 4 \, A_\Delta \left ( 
{\partial^2 \over \partial a \; \partial b} \; + \; 
{\partial^2 \over \partial b \; \partial c} \; + \; 
{\partial^2 \over \partial c \; \partial a} \right ) \; .
\label{eq:lapl_2d}
\eeq
Note that the curvature term involves three deficit angles
$\delta_i$, associated with the three vertices of a triangle.
Now, Eq.~(\ref{eq:wd_2d}) applies to a single given triangle, with one
equation to be satisfied at each triangle on the lattice. 
One can also construct the total Hamiltonian by
simply summing over all triangles, which leads to
\beq
\left \{ - \, ( 16 \pi \, G)^2 \, \sum_{\Delta} \, {\bf \Delta } (l^2 ) 
\, + \, 2 \, \lambda \sum_{\Delta} A_{\Delta}
\, - \, \kappa \, \sum_{\Delta} \sum_{ i \subset \Delta } \delta_i 
\right \} \, \Psi = 0 \; .
\label{eq:ham_tot}
\eeq
Summing over all triangles $(\Delta)$ is different from summing over
all lattice sites $(i)$, and the
above equation is equivalent to 
\beq
\left \{ - \, ( 16 \pi \, G)^2 \,  \sum_{\Delta} {\bf \Delta } (l^2) \, + \,
2 \, \lambda \sum_{\Delta} A_\Delta \, - \,
\kappa \, q \, \sum_i \delta_i \right \} \, \Psi = 0 \; ,
\eeq
where $q$ is the lattice coordination number and is determined
by how the lattice is put together (which vertices are neighbors to
each other, or, equivalently, by the so-called incidence matrix).
Here, $q$ is the number of neighboring simplexes that share a given
hinge (vertex).
For a flat triangular lattice $q=6$, whereas
for a tetrahedron, octahedron, and icosahedron, one has $q=3,\, 4,\, 5$, respectively.
For proper normalization in Eq.~(\ref{eq:ham_tot})
one requires 
\beq
\int d^2 x \, \sqrt{g} \;\;\;  \longleftrightarrow \sum_{\Delta} A_\Delta
\eeq
as well as 
\beq
\half \,  \int  \, d^2 x \, \sqrt{g} \; R  
\, \longleftrightarrow   \sum_i \delta_i \; .
\eeq
This last correspondence allows one to fix the overall normalization of the curvature term
\beq
\kappa \, \equiv \, 2 \, n_{\sigma \, h} \, = \, { 2 \over q } \; ,
\label{eq:kappa}
\eeq
which then determines the relative weight of the local volume and curvature terms.

\vskip 40pt

\section{Choice of Coupling Constants}

\label{sec:choiceofunits}

As in the Euclidean lattice theory of gravity, we will find
it convenient here to factor out an overall irrelevant length
scale from the problem and set the (unscaled) cosmological constant equal to one 
as was done in \cite{hw84}.
Indeed recall that the Euclidean path integral weight always contains a factor 
$ P(V) \propto \exp (- \lambda_0 V) $, where 
$V = \int \sqrt{g} $ is the total volume on the lattice, and $ \lambda_0 $ 
is the unscaled cosmological constant.
The choice $\lambda_0=1$ then fixes this overall scale once and for all.
Since $\lambda_0 = 2 \lambda / 16 \pi G $, one then has 
$\lambda = 8 \pi G  $ in this system of units.
In the following we will also find it convenient to introduce a scaled coupling 
$\tilde \lambda$ defined as
\beq
\tilde \lambda \; \equiv \; { \lambda \over 2 } \left ( { 1 \over 16 \pi G } \right )^2
\label{eq:tilde}
\eeq
so that for $\lambda_0=1 $ (in units of the $UV$ cutoff or, equivalently, 
in units of the fundamental lattice spacing) one has 
$\tilde \lambda = 1/ 64 \pi G $.
One can now rewrite the Wheeler-DeWitt equation so that the kinetic term (the term
involving the Laplacian) has a unit coefficient and write Eq.~(\ref{eq:wd_1a}) as
\beq
\left \{  -  \, {\bf \Delta } \, + \, 
{2 \lambda \over  \left(16 \pi G \right)^2 } \, \sqrt{g} \, - \, 
{1 \over  \left(16 \pi G \right)^2 }  \, \sqrt{g} \, R \, \right \} \, \Psi = 0 \;.
\label{eq:wd_c}
\eeq
Note that in the extreme strong coupling limit ($G \rightarrow \infty$)
the kinetic term is the dominant one, followed by the volume 
(cosmological constant) term (using the facts about $\tilde \lambda$ given
above) and, finally, by the curvature term.
Consequently, at least in a first approximation, the curvature $R$ term can be
neglected compared to the other two terms in this limit.

Two further notational simplifications will be done in the following.
The first one is introduced in order to avoid lots of factors of 
$16 \pi$ in many of the subsequent formulas.
Consequently, from now on we shall write $G$ as a shorthand for $16 \pi\, G$,
\beq
16 \pi \, G \longrightarrow G \; .
\label{eq:shortG}
\eeq
In this notation one then has $\lambda = G / 2 $ and 
$\tilde \lambda = 1/ 4 G $.
The above notational choices then lead to a much more streamlined representation of
the Wheeler-DeWitt equation,
\beq
\left \{  -  \, {\bf \Delta } \, + \, 
{ 1 \over  G } \, \sqrt{g} \, - \, 
{ 1 \over G^2 }  \, \sqrt{g} \, R \, \right \} \, \Psi = 0 \;.
\label{eq:wd_c1}
\eeq
A second notational choice will be dictated later on by the structure
of the wave function solutions, which will commonly involve factors of $\sqrt{G}$.
For this reason we will now define the new coupling $g$ as
\beq
g \; \equiv \; \sqrt{G} \; ,
\label{eq:gdef}
\eeq
so that $\tilde \lambda = 4 / g^2 $ 
(the latter $g$ should not be confused with the square root of the
determinant of the metric).

Later on it will be convenient to define a parameter $\beta$ for
the triangulations of the sphere, defined as
\beq
\beta \; \equiv \; { 2 \, \pi  \over \sqrt{\tilde{\lambda}} \, G^2} \; .
\label{eq:beta}
\eeq
Factors of $2 \pi $ arise here because we are looking
at various triangulations of the two-sphere.
More generally, for a two-dimensional closed manifold with arbitrary
topology, one has by the Gauss-Bonnet theorem
\beq
\int d^2 x \, \sqrt{g} \, R \; = \; 4 \pi \, \chi
\label{eq:euler}
\eeq
with $\chi $ as the Euler characteristic of the manifold.
The latter is related to the genus $g$ (the number of handles)
via $\chi = 2 - 2 g $
(note that for a discrete manifold in two dimensions one has the 
equivalent form due to Euler $\chi = N_0 - N_1 + N_2 $, where $N_i$
denotes the number of simplices of dimension $i$).
Thus for a general two-dimensional manifold we will define
\beq
\beta \; = \; { \pi \, \chi \over \sqrt{\tilde{\lambda}} \, G^2} \;.
\label{eq:beta_chi}
\eeq
Equivalently, using
\beq
\sqrt{\tilde{\lambda}} \, G^2 \, = \, 
{ 1 \over 2 \sqrt{G} } \cdot G^2 \, = \, \half \, G^{3/2}
\eeq
and then making use of the coupling $g$, one has simply
\beq
\beta \; = \; { 4 \pi \over g^3 } \; 
\label{eq:beta_g}
\eeq
for the sphere, and in the more general case
\beq
\beta \; = \;  { 2 \pi \, \chi \over g^3} \; .
\label{eq:beta_g_chi}
\eeq

\vskip 40pt

\section{Outline of the General Method of Solution}
\label{sec:generalcase}

It should be clear from the previous discussion that
in the strong coupling limit (large $G$) one can, at least at first,
neglect the curvature term, which can then be included at
a later stage.
This simplifies the problem quite a bit, as it is the curvature 
term that introduces complicated interactions between neighboring
simplices
(this is evident from the lattice Wheeler-DeWitt equation of
Eq.~(\ref{eq:wd_latt1}), where the deficit angles enter the curvature term only).

The general procedure for finding a solution will be as follows.
First a solution will be found for {\it{equilateral}} edge lengths 
$s$.
Later this solution will be extended to
determine whether it is consistent to higher order in the weak field
expansion.
Consequently we shall write for the squared edge lengths 
\beq
l_{ij}^2 \; = \; s \left( 1 \, + \, \epsilon \, h_{ij} \right) \; ,
\label{eq:wfe}
\eeq
with $\epsilon$ a small expansion parameter.
Therefore, for example, in Eq.~(\ref{eq:lapl_2d}) one has
$a = s (1 + \epsilon h_a)$, $b = s (1 + \epsilon h_b)$ and $c = s (1 +
\epsilon h_c)$.
The resulting solution for the wave function will then be given by a 
suitable power series in the $h$ variables.
Nevertheless, in some rare cases (such as the single-triangle case described
below or the single tetrahedron in $3+1$ dimensions \cite{hw11}),
one is lucky enough to find immediately an exact solution,
without having to rely in any way on the weak field expansion.

To lowest order in $h$, a solution to the Wheeler-DeWitt equation
is readily found using the standard power series (or Frobenius)
method, appropriate for the study of quantum mechanical wave equations.
In this method one first obtains the correct asymptotic behavior
of the solution for small and large arguments and later constructs
a full solution by writing the remainder as a power series or
polynomial in the relevant variable.
Of some importance in the following is the correct determination of
the wave functional $\Psi$ for small and large areas (small and large 
$s$), and to what extent the resulting wave function can be expressed in
terms of invariants such as areas and curvatures, or powers thereof.

In the following we will see that the natural variable for displaying 
results is the scaled total area $x$, defined as
\beq
x \; \equiv \; 2 \sqrt{\tilde{\lambda}}  \, A_{tot} \; = \; 
2 \sqrt{\tilde{\lambda}}  \, \sum_{\Delta} \, A_{\Delta} \; .
\label{eq:xdef}
\eeq
We will look at a variety of two-dimensional lattices, including
the regular triangulations of the two-sphere given by the
tetrahedron, octahedron and icosahedron, as well as the case
of a triangulated torus with coordination number six.
In the equilateral case the natural variable for displaying the
results is then
\beq
x \; = \; 2 \, \sqrt{\tilde{\lambda}}  \, A_{tot} \; = \; 2 \, N_{\Delta}
\, \sqrt{\tilde{\lambda}}  \, A_{\Delta} \; .
\eeq
Later on we will be interested in taking the 
infinite volume limit, defined in the usual way as
\bea
N_{\Delta} & \rightarrow & \infty \; ,
\nonumber \\
A_{tot} & \rightarrow & \infty \; ,
\nonumber \\
{ A_{tot} \over N_{\Delta} } & \rightarrow & {\rm const.} \; .
\label{eq:infvol}
\eea
It follows that this last ratio can be used to define a fundamental lattice
spacing $l_0$, for example via 
$ A_{tot} / N_{\Delta}  = A_{\Delta} = \sqrt{3} \, l_0^2 / 4 $.

The full solution of the quantum mechanical problem
will, in general, require that the wave functions be
properly normalized, as in Eq.~(\ref{eq:norm}).
This will introduce at some stage wave function normalization factors
${\mathcal{N}}$ and $ \tilde{ {\mathcal{N}} }$,
which will be fixed by the standard rules of quantum mechanics.
If the wave function depends on the total area only, then the relevant
requirement becomes
\beq
\int_0^\infty  d A_{tot}  \; | \, \Psi (A_{tot}) \, |^2 
\; \equiv \; { 1 \over 2 \, \sqrt{\tilde{\lambda}} } \, 
\int_0^\infty  d x \; | \, \Psi (x) \, |^2 \; = \; 1 \; .
\label{eq:normaliz}
\eeq
As in nonrelativistic quantum mechanics, two solutions are
expected, only one of which will be regular as the origin
and thus satisfy the wave function normalizability
requirement.

At this point it will be necessary to discuss each lattice separately
in some detail.
For each lattice geometry, we will break down the presentation into four
separate items:

\noindent
{\bf{(a) Equilateral Case in the Strong Coupling Limit $(\epsilon = 0 )$}}.
This subsection will find a solution in the extreme strong coupling limit (large
$G$), without curvature term in the Wheeler-DeWitt equation.
The solution will not rely on the weak field expansion, and the
results will be exact to zeroth order in the weak field expansion
of Eq.~(\ref{eq:wfe}).
In this case the simplices are all taken to be equilateral, and
the lattice edge lengths fluctuate together.

\noindent
{\bf{(b) Equilateral Case with Curvature Term $(\epsilon = 0 )$}}.
Next, the curvature term is included. 
The solution again will not rely on the weak field expansion, and
all the triangles will be taken to be equilateral.
The resulting solution will, therefore, be valid again (and exact) to 
zeroth order in the $\epsilon$ expansion parameter of Eq.~(\ref{eq:wfe}).

\noindent
{\bf{(c) Large Area in the Strong Coupling Limit $(\epsilon \ne 0 )$}}.
In this case we will look at nonzero local fluctuations 
in Eq.~(\ref{eq:wfe}).
The method of solution will now rely on the weak field 
expansion for large areas (large $s$), but nevertheless it will 
turn out that an exact solution can be found in this case.
The resulting answer will be found to be correct 
to arbitrarily large order ${\mathcal{O}}(\epsilon^n )$,
with $n$ a positive integer.

\noindent
{\bf{(d) Small Area in the Strong Coupling Limit $(\epsilon \ne 0 )$}}.
Finally we will look at the case of nonzero fluctuations 
[$ \epsilon \ne 0 $ in Eq.~(\ref{eq:wfe})] in the
limit of small areas (small $s$).
In this limit we will find that, in general, the solution can be written
entirely in terms of invariants involving total areas and
curvatures only up to order 
${\mathcal{O}} ( \epsilon ) $ or ${\mathcal{O}} ( \epsilon^2 ) $,
depending on whether a further symmetrization of the problem
is performed or not.

If the reader is not interested in the details for each lattice,
he can skip the next few subsections and go directly to the
summary presented in Sec.~(\ref{sec:summary}).

\vskip 40pt

\subsection{Single Triangle Case}
\label{sec:singletriangle}

From Eq.~(\ref{eq:wd_2d}) the Wheeler-DeWitt equation for a single
triangle reads 
\beq
\left  \{  \, (16 \pi G)^2 \; 4 A_\Delta \; \left ( {\partial^2 \over \partial a
\; \partial b} \; + \; {\partial^2 \over \partial b \; \partial c} \;
+ \; {\partial^2 \over \partial c \; \partial a} \right )
\; + \; 2 \lambda \, A_\Delta \right \} \, \Psi ( \, a,b,c \, ) \; = \; 0,
\eeq
where 
$a,b,c$ 
are the three squared edge lengths for the given triangle, and 
$A_\Delta$ 
is the area of the same triangle.
Note that for a single triangle there can be no curvature term.
Equivalently one needs to solve
\beq
\left  \{  
{\partial^2 \over \partial a
\; \partial b} \; + \; {\partial^2 \over \partial b \; \partial c} \;
+ \; {\partial^2 \over \partial c \; \partial a}
\; + \; \tilde \lambda 
\right \} \Psi ( \, a,b,c \, ) \; = \; 0 \; .
\eeq
If one sets 
\beq
\Psi[ \, a,b,c \, ] \; = \; \Phi[ \, A_\Delta \, ] \; ,
\eeq
then one finds the following equivalent differential equation
\beq
A_\Delta \; { d^2 \Phi \over dA_\Delta ^2 } \; + \; 2 \; { d \Phi \over dA_\Delta }  
\; + \; 16 \; \tilde \lambda \; A_\Delta \; \Phi \; = \; 0 \; .
\label{eq:tria_eq}
\eeq
For a single triangle the total area equals the area of the single triangle,
$A_{tot} = A_{\Delta}$.
Here it will be convenient to define 
\beq
x \, = \, 4 \sqrt{\tilde{\lambda}}  \, A_{tot} \; \equiv \; 
4 \sqrt{\tilde{\lambda}}  \, A_{\Delta}
\eeq
so that the solution will be function of this variable only. 
Note that in this case, and in this case only, we will deviate from
the general definition of the variable $x$ given in Eq.~(\ref{eq:xdef}).
One can then write the solution to Eq.~(\ref{eq:tria_eq}) in
the form
\beq
\Psi(x) \; = \; 
\mathcal{N} \, { J_n (x) \over x^n } 
\eeq
with 
\beq
n \; = \; {1 \over 2} \; 
\eeq
so that
\beq
\Psi(x) \; = \; 
{\mathcal{N}} \,  
{ J_{1/2} \left ( \, 4 \sqrt{\tilde{\lambda}} \, A_{tot}  \right) 
\over \left( 4 \sqrt{\tilde{\lambda}} \, A_{tot} \right)^{1/2}  } \; .
\eeq
The wave function normalization constant is given here by
\beq
{\mathcal{N}} \; = \; 2 \, \tilde{\lambda}^{1/4} \; .
\eeq
Note that the above solution is exact, and did not require in any way
the weak field expansion.
Two alternate forms of the wave function are
\bea
\Psi ( A_{tot} ) & = &
\mathcal{N}  \; {\sin\left(4 \sqrt{\tilde{\lambda}} A_{tot}  \right) 
\over 2 \, \sqrt{2 \pi} \, \sqrt{\tilde{\lambda}} \, A_{tot}} \nonumber \\
& = &
\mathcal{N} \; \sqrt{ 2 \over \pi } \, 
\exp \left ( - \, 4 \, i \, \sqrt{\tilde{\lambda}} \, A_{tot} \right ) 
\, _1 F_1 \left ( 1, 2,  8 \, i \, \sqrt{\tilde{\lambda}} \, A_{tot}
\right ) \; .
\eea
Here $_1 F_1 (a,b,z) $ is the confluent hypergeometric functions of
the first kind.
The usefulness of the latter representation will become clearer later,
when other lattices are considered and the curvature term
is included.
Expanding the solution for small area one obtains
\beq
\Psi (x) \; = \;   
\mathcal{N} \; \sqrt{ 2 \over \pi } \, 
\left[1- {x^2 \over 6} + {x^4 \over 120} + {\mathcal{O}}\left(x^6\right) \right]
\eeq
which shows that it is indeed nonsingular and, thus, normalizable.

In the limit of large areas, a solution to the original differential
equation is given either by the asymptotic
behavior of the above Bessel (here sine) function (J), the same limiting
behavior for the corresponding
Bessel function Y, or by the two corresponding Hankel functions (H).
\beq
\Psi \; \mathrel{\mathop\sim_{ x \; \rightarrow \; \infty } } \; 
{ 1 \over x } \; \exp \left ( \pm \, i \,x \right ) 
\; \sim  \; { 1 \over A_{tot} }
\; \exp \left ( \pm \, 4 \, i \, \sqrt{ \tilde{\lambda} } \, A_{tot} \right )  \; .
\eeq
Nevertheless among those four solutions, only one is regular and,
therefore, physically acceptable.

The calculation for a single triangle can be regarded as
a useful starting point, and a basic stepping stone, for the 
strong coupling expansion in  $1/G$.
It shows the physical characteristics of the wave function 
solution deep in the strong coupling regime:
for $ G \rightarrow \infty $ 
the coupling term between different simplices, which is caused mainly by the
curvature term,
disappears and one ends up with a completely decoupled problem, 
where the edge lengths in nonadjacent simplices fluctuate independently.

\vskip 40pt

\subsection{Tetrahedron}
\label{sec:tetrahedron}

In the case of the tetrahedron one has 4 triangles, 6 edges, and 4
vertices, and 3 neighboring triangles for each vertex.
Let us discuss again, here, the various cases individually.

\noindent
{\bf{(a) Equilateral Case in the Strong Coupling Limit $\left(\epsilon = 0\right)$}}

We first look at the case when $\epsilon = 0 $ in Eq.~(\ref{eq:wfe}), 
deep in the strong coupling region and without the curvature term.

Following Eq.~(\ref{eq:xdef}) we define the scaled area variable as
\beq
x \; = \; 2 \sqrt{\tilde{\lambda}}  \, A_{tot} \; = \; 
4 \times 2 \sqrt{\tilde{\lambda}}  \, A_{\Delta}
\eeq
and the solution will be found later to be a function of this variable only.
For equilateral triangles the wave function $\Psi $ needs to satisfy
\beq
\Psi'' \, + \, { 2 \over x } \, \Psi' \, + \, \Psi \, = \, 0 \; .
\label{eq:tet_eq}
\eeq
The correct solution can be written in the form
\beq
\Psi(x) \; = \; \mathcal{N} \, {J_n (x) \over x^n} 
\eeq
with
\beq
n \; = \; {1 \over 2}
\eeq
so that
\beq
\Psi(x) \; = \; {\mathcal{N}} \;
{ J_{1/2} \left( 2 \sqrt{\tilde{\lambda}} \, A_{tot} \right) 
\over    \left( 2 \sqrt{\tilde{\lambda}} \, A_{tot} \right)^{1/2}  } \; .
\eeq
The wave function normalization constant is given by
\beq
{\mathcal{N}} \; =  \; \sqrt{2} \;  \tilde{\lambda} ^{1\over 4} \; .
\eeq
Below are two equivalent forms of the same wave function
\bea
\Psi ( A_{tot} ) & = &
\mathcal{N}  \; {\sin \left( {2 \sqrt{\tilde{\lambda}} \, A_{tot} } \right) 
\over \sqrt{2 \pi} \, \sqrt{\tilde{\lambda}}  \, A_{tot}} 
\nonumber \\
& = &
\mathcal{N} \; \sqrt{ 2 \over \pi } \,
\exp \left ( - \,  2 \, i \, \sqrt{\tilde{\lambda}} \, A_{tot} \right ) \, 
_1F_1\left ( 1, 2,  4 \, i \, \sqrt{\tilde{\lambda}} \, A_{tot} \right ) 
\eea
for the equilateral case.
In the limit of small area one obtains
\beq
\Psi \; = \;  \mathcal{N} \; 
\sqrt{ 2 \over \pi } \,
\left[1- {x^2 \over 6} + {x^4 \over 120 } + {\mathcal{O}}\left(x^6\right) \right]
\eeq
which again confirms that the wave function is regular at the origin.
Since one is solving a second-order linear differential equation, 
one expects two solutions; here, one is singular and the other one is
not, as is often the case in quantum mechanics.
For the geometry of the tetrahedron, one solution can be 
written in terms of Bessel functions of the first kind ($J$)
\beq
{J_{1/2}(x) \over \sqrt{x}} \,  = \, \sqrt{ 2 \over \pi } \, { \sin x
  \over x } \; .
\eeq
The Bessel function of the second kind ($Y$) also satisfies
the same differential equation, but since
\beq
{Y_{1/2} (x) \over \sqrt{x}} \, = \, - \sqrt{ 2 \over \pi } \, { \cos x  \over x }
\eeq
this second solution is not normalizable, it is therefore discarded
on physical grounds.
We shall see below that the same behavior at small $x$ holds also
for the nonzero curvature term.
Note that both of the above solutions are real.
\footnote{
There are also linear combinations of Bessel functions 
which give complex Hankel ($H$) functions.
These satisfy the Wheeler-DeWitt equation as well;
however, they are not physically acceptable since both are singular at
the origin.}

\noindent
{\bf{(b) Equilateral Case with Curvature Term $\left(\epsilon = 0\right)$}}

Next we include the effects of the curvature term.
To zeroth order in weak field expansions, when
all edges fluctuate in unison, one now needs to solve the
ordinary differential equation
\beq
\Psi'' \, + \, { 2 \over x } \, \Psi' \, - \,
{ 2 \beta \over x } \, \Psi \, + \, \Psi \, = \, 0 \; ,
\label{eq:tet_eq_R}
\eeq
with $ \beta = 2 \pi / \sqrt{\tilde{\lambda}} \, G^2 $ as in Eq.~(\ref{eq:beta}).
Since the deficit angle $\delta = \pi$ at each vertex, the curvature
contribution for each triangle is $ \kappa \cdot \pi \cdot 3 $.
In this case one has, therefore,
\beq
\kappa_{tetra} = 2 \cdot {1 \over 3}
\eeq
and, therefore, the solution is given by
\bea
\Psi & \simeq &
\exp \left ( - \, 2 \, i \, \sqrt{ \tilde{\lambda} } \, A_{tot} \right) \; 
_1F_1 \left ( 1 - i \, 
{ 3 \, \pi \, \kappa_{tetra} \over G^2 \, \sqrt{ \tilde{\lambda} } }, \; 
2,\; 4 \, i \, \sqrt{ \tilde{\lambda} } \, A_{tot} \right )  
\nonumber \\
& = &
\exp \left( - \, 2 \, i \, \sqrt{ \tilde{\lambda} } \, 
A_{tot} \right) \;
_1F_1 \left ( 1 - i \, 
{ 2 \, \pi \, \over G^2 \, \sqrt{ \tilde{\lambda} } },
\; 2,\; 4 \, i \, \sqrt{ \tilde{\lambda} } \, A_{tot} \right )
\eea
in the equilateral case, up to an overall normalization factor.
Note that in this case one had to include a factor of
$A_{tot}/ (4 A_\Delta ) $ (which in the tetrahedron case equals one)
in the imaginary part of the first argument of $_1F_1$.

\noindent
{\bf{(c) Large Area in the Strong Coupling Limit $\left(\epsilon \ne 0\right)$}}

Next we look at the case $ \epsilon \ne 0 $ in Eq.~(\ref{eq:wfe}).
In the limit of large areas one finds that the two independent solutions reduce to 
\beq
\Psi 
\; \mathrel{\mathop\sim_{ x \; \rightarrow \; \infty } } \; 
\exp\left ( \pm \, i \,x \right ) \; \sim \; \exp \left 
( \pm \, 2 \, i \, \sqrt{\tilde{\lambda}} \, A_{tot} \right )  \; 
\label{eq:wf_tet_s}
\eeq
to all orders in $\epsilon$.
To show this, one sets $ \Psi = e^{\alpha \, A_{tot}} $, where $ A_{tot} $ 
is a sum of the four triangle areas that make up the tetrahedron, and 
then expands the edge lengths in the usual way according to
Eq.~(\ref{eq:wfe}), 
by setting $ a = s ( 1 + \epsilon \, h_a ) $, etc.
Here we are interested specifically in the limit when $ s $ is large and $ \epsilon $
is small.
One then finds that the rhs of the lattice Wheeler-DeWitt 
equation is given to $ {\mathcal{O}} ( \epsilon^n ) $ by 
\beq
{e^{\alpha \, \sqrt{3} \, s } \over 4 } \, 
{1 \over 2^n \, {\sqrt{3} \, ^n} \, n! } \, 
\alpha^n \, \left( \alpha^2 + 4 \,  {\tilde{\lambda}} \right) \, 
\epsilon^n \, s^n \, \left( \sum h \right)^n + \cdots \;\; .
\eeq
One concludes that in this limit it is sufficient to have
\beq
\alpha^2 + 4 \,  {\tilde{\lambda}}  \; = \; 0 ,
\eeq
or $\alpha = \pm \, 2 \, i \,  \sqrt{\tilde{\lambda}}$, to obtain
an exact solution in the limit $n \rightarrow \infty $.
Note that in the strong coupling limit the two independent wave function solutions in
Eq.~(\ref{eq:wf_tet_s}) completely factorize as a product 
of single-triangle contributions.

\noindent
{\bf{(d) Small Area in the Strong Coupling Limit $\left(\epsilon \ne 0\right)$}}

In the limit of small area, we have shown before that
the solution reduces to a constant in the equilateral case 
[$ {\mathcal{O}}\left(\epsilon^0 \right) $] for
small $x$ or small areas.
Beyond the equilateral case one can write a general ansatz 
for the wave function in terms of geometric invariants
\beq
\Psi = \left( \prod_\Delta A_\Delta \right)^{\gamma_0} \, \left[ 1 \, + \,
\gamma_2 \, \left( \sum_\Delta A_\Delta \right)^2 \, + \, 
\gamma_4 \, \left( \sum_\Delta A_\Delta \right)^4 \, + \, \cdots
\right] \; ,
\label{eq:ansatz}
\eeq
and then expand the solution in $\epsilon$ for small $s$.
To zeroth order in $\epsilon$ we had the solution
$ \Psi \sim J_n (x) / x^n  $ with 
$ x = 2  \, \sqrt{\tilde{\lambda}} \, A_{tot}$ and $n = 1/2 $.
This gives in Eq.~(\ref{eq:ansatz}) $\gamma_0 =0$,
$\gamma_2 = - \, {2 \over 3} \,  {\tilde{\lambda}} $ and
$\gamma_4 = {2 \over 15} \, {\tilde{\lambda}} ^2 $.
To linear order [${\mathcal{O}}\left(\epsilon \right)$]
one finds, though, that terms appear which cannot be
expressed in the form of  Eq.~(\ref{eq:ansatz}).
But one also finds that, while these terms are nonzero if one uses the Hamiltonian 
density (the Hamiltonian contribution from just a single triangle),
if one uses the sum of such triangle Hamiltonians, then the 
resulting solution is symmetrized, and the corrections to
Eq.~(\ref{eq:ansatz}) are found to be of order 
${\mathcal{O}}\left(\epsilon^2 \right)$.
Then the wave function for small area is of the form
\beq
\Psi \sim \; 
1 \, - \, \twoth \, {\tilde{\lambda}} \, A_{tot}^2 \, + \, 
{\textstyle {2 \over 15} \displaystyle}
\, {\tilde{\lambda}}^2 \, A_{tot}^4 \, + \, \dots 
\eeq
up to terms ${\mathcal{O}} ( \epsilon^2 )$.

\vskip 40pt

\subsection{Octahedron}
\label{sec:octahedron}

The discussion of the octahedron proceeds in a way that is similar to
what was done before for the tetrahedron.
In the case of the octahedron one has 8 triangles, 12 edges
and 6 vertices,
with 4 neighboring triangles per vertex.
Again we will now discuss the various cases individually.

\noindent
{\bf{(a) Equilateral Case in the Strong Coupling Limit $\left(\epsilon = 0\right)$}}

Again we look first at the case $\epsilon = 0 $ in Eq.~(\ref{eq:wfe}), 
deep in the strong coupling region and without the curvature term.
Following Eq.~(\ref{eq:xdef}) we define the scaled area variable as
\beq
x  = 2 \, \sqrt{\tilde{\lambda}}  \, A_{tot} \; = \; 
8 \times 2 \, \sqrt{\tilde{\lambda}}  \, A_{\Delta}
\eeq
and it is found that the solution is a function of this variable only.
For equilateral triangles the wave function $\Psi $ needs to satisfy
\beq
\Psi'' \, + \, { 4 \over x } \, \Psi' \, + \, \Psi \, = \, 0 \; .
\label{eq:oct_eq}
\eeq
The correct solution can be written in the form
\beq
\Psi(x) 
\; = \; \mathcal{N} \; {J_n (x) \over x^n} 
\eeq
with
\beq
n \; = \; {3 \over 2}
\eeq
so that
\beq
\Psi(x) \; = \; \mathcal{N} \; 
{ J_{3/2} \left( 2 \,  \sqrt{\tilde{\lambda}} A_{tot} \right) \over 
\left( 2 \,  \sqrt{\tilde{\lambda}} A_{tot} \right)^{3/2}} \; .
\eeq
The wave function normalization factor is given by
\beq
\mathcal{N} \; = \; \sqrt{15} \, \tilde{\lambda}^{1/4} \; .
\eeq
Equivalent forms of the above wave function are
\bea
\Psi ( A_{tot} ) & = &
\mathcal{N} \; { 1 \over 2^{3/2} \; \Gamma\left({5 \over 2} \right) } \, 
\exp \left ( - \, 2 \, i \, \sqrt{\tilde{\lambda}} \, A_{tot}  \right ) \;
_1F_1 \left ( 2,4, 4 \, i \, \sqrt{\tilde{\lambda}} \, A_{tot}  \right) 
\nonumber \\
&=&
\mathcal{N} \; 
\left [ -\, 
{ \cos \left( 2 \, \sqrt{\tilde{\lambda}} \, A_{tot} \right) 
\over 2 \, \sqrt{2 \pi} \,  \tilde{\lambda} \, A_{tot}^2 } + 
{ \sin \left( 2 \, \sqrt{\tilde{\lambda}} \, A_{tot} \right)  
\over 4 \,  \sqrt{2 \pi}  \, \tilde{\lambda}^{3/2}  \, A_{tot}^3} \,
\right ] \; .
\eea
These can be expanded for small $A_{tot}$ or small $x$ to give
\beq
\Psi  \; = \;   
\mathcal{N} \; { \sqrt{2} \over 3 \, \sqrt{\pi} } \;  
\left[ 1- {x^2 \over 10} + {x^4 \over 280} + {\mathcal{O}}(x^6) \right] \; .
\eeq
We note here again that both Bessel functions of the first ($J$) and
second ($Y$) kind, in principle, give solutions for this case, 
as well as the two corresponding Hankel ($H$) functions.
Nevertheless, only the solution associated with the Bessel $J$ function
is regular near the origin.

\noindent
{\bf{(b) Equilateral Case with Curvature Term $\left(\epsilon = 0\right)$}}

Next, we include the effects of the curvature term.
Since here the deficit angle $\delta = 2 \pi / 3 $ at each vertex, the curvature
contribution for each equilateral triangle is 
$ \kappa \cdot { 2 \, \pi \over 3 } \cdot 3 = 2 \pi \, \kappa $.
For the octahedron one has in Eq.~(\ref{eq:kappa})
\beq
\kappa_{octa} \; = \; 2 \cdot {1 \over 4} \; .
\eeq
With the curvature term one finds
\bea
\Psi ( A_{tot} ) 
& \simeq & 
\exp \left ( - \, 2 \, i \, \sqrt{\tilde{\lambda}} \, 
A_{tot} \right ) \;  
_1F_1 \left ( 2 - i \, {4 \pi \, \kappa_{octa} 
\over \sqrt{\tilde{\lambda}} \, G^2},\; 4,\; 4 \, i \, 
\sqrt{\tilde{\lambda}} \, A_{tot} \right ) 
\nonumber \\
& = & 
\exp \left ( - \, 2 \, i \, \sqrt{\tilde{\lambda}} \, 
A_{tot} \right ) \; 
_1F_1 \left ( 2 - i \, {2 \, \pi \over \sqrt{\tilde{\lambda}} \, 
G^2},\; 4,\; 4 \, i \, \sqrt{\tilde{\lambda}} \, A_{tot}  \right )
\; .
\eea
Note that in this case one had to include a factor 
$A_{tot}/( 4 A_\Delta) $, which in the octahedron case equals two.

\noindent
{\bf{(c) Large Area in the Strong Coupling Limit $\left(\epsilon \ne 0\right)$}}

In the limit of large areas the two independent solutions reduce to
\beq
\Psi 
\; \mathrel{\mathop\sim_{ x \; \rightarrow \; \infty } } \; 
\exp\left ( \pm \, i \,x \right )
\; \sim \; \exp \left ( \pm \, 2 \, i \, \sqrt{\tilde{\lambda}}  \, A_{tot}
\right ) \; 
\eeq
to all orders in $\epsilon$.
In other words, to $ {\mathcal{O}}\left(\epsilon^n\right) $
with $n \rightarrow \infty $, as for the tetrahedron case.
Note also that in the strong coupling limit the two independent
wave function solutions again completely factorize as a 
product of single-triangle contributions.

\noindent
{\bf{(d) Small Area in the Strong Coupling Limit $\left(\epsilon \ne 0\right)$}}

In the limit of small area, the solution approaches 
a constant in the equilateral case.
To go beyond the equilateral case, one can write again a general ansatz 
for the wave function, written in terms of geometric invariants
as in Eq.~(\ref{eq:ansatz}).
Then the solution can be expanded in $\epsilon$ for small $s$.
To zeroth order in $\epsilon$, the solution is
$ \Psi \sim J_n (x) / x^n  $ with $n = 3/2 $.
This gives in Eq.~(\ref{eq:ansatz})
$\gamma_0 =0$, $ \gamma_2 = - \, {2 \over 5} \, {\tilde{\lambda}}$
and $ \gamma_4  = {2 \over 35} \,  {\tilde{\lambda}}^2$.
However, to linear order [${\mathcal{O}}\left(\epsilon \right)$] 
one finds again that linear terms in $h$ appear which cannot be
expressed in the form of  Eq.~(\ref{eq:ansatz}).
But one also finds that while these terms are nonzero if one uses the Hamiltonian 
density (the Hamiltonian contribution from just a single triangle),
if one uses the sum of such triangle Hamiltonians then the 
resulting solution is symmetrized, and the corrections to
Eq.~(\ref{eq:ansatz}) are found to be of order 
${\mathcal{O}}\left(\epsilon^2 \right)$.
Then the wave function for small area is of the form
\beq
\Psi \, \simeq \,  
1 \, - \, {\textstyle {2 \over 5} \displaystyle}
{\tilde{\lambda}} \, A_{tot}^2  
\, + \, {\textstyle {2 \over 35} \displaystyle}
\, {\tilde{\lambda}}^2 \, A_{tot}^4 \, + \, \dots
\eeq
up to terms of ${\mathcal{O}} ( \epsilon )$.

\vskip 40pt

\subsection{Icosahedron}
\label{sec:icosahedron}

The discussion of the icosahedron proceeds in a way that is similar to
what was done before for the other regular triangulations.
Here one has 20 triangles, 30 edges and 12 vertices,
with 5 neighboring triangles per vertex.
Let us again discuss the various cases individually.

\noindent
{\bf{(a) Equilateral Case in the Strong Coupling Limit $\left(\epsilon = 0\right)$}}

Again we look first at the case $\epsilon = 0 $ in Eq.~(\ref{eq:wfe}), 
deep in the strong coupling region and without curvature term.
Following Eq.~(\ref{eq:xdef}) we define the scaled area variable as
\beq
x \; = \; 2 \, \sqrt{\tilde{\lambda}}  \; A_{tot} \; 
\equiv \; 20 \times 2 \, \sqrt{\tilde{\lambda}}  \; A_{\Delta}
\eeq
and a solution is found which is a function of this variable only.
For equilateral triangles the wave function $\Psi $ needs to satisfy
\beq
\Psi'' \, + \, { 10 \over x } \, \Psi' \, + \, \Psi \, = \, 0 \; .
\label{eq:ico_eq}
\eeq
A solution can then be found of the form
\beq
\Psi(x) \, = \, \mathcal{N} \; {J_n (x) \over x^n} 
\eeq
with 
\beq
n \; = \;  {9 \over 2}
\eeq
so that
\beq
\Psi(x) \, = \, \mathcal{N} \; 
{ J_{9/2}\left( 2 \, \sqrt{\tilde{\lambda}} \, A_{tot}  \right)
\over   \left( 2 \, \sqrt{\tilde{\lambda}} \, A_{tot} \right)^{9/2} } \; .
\eeq
The wave function normalization factor is given by
\beq
\mathcal{N} \; = \; 9 \, \sqrt{12155} \; \tilde{\lambda}^{1/4} \; .
\eeq
Below is an equivalent form of the same solution
\beq
\Psi ( A_{tot} ) \; = \; 
\mathcal{N} \;  {1 \over 2^{9/2} \; \Gamma\left( {11 \over 2} \right)}
\;
\exp \left ( - \, 2 \, i \, \sqrt{\tilde{\lambda}} \, A_{tot} \right ) \;
_1F_1 \left ( 5, \, 10,  \, 4 \, i \, \sqrt{\tilde{\lambda}} \,
A_{tot} \right ) \; .
\eeq
For small area $A_{tot}$ or small $x$, one obtains
\beq
\Psi \; = \; \mathcal{N} \;  {1 \over 2^{9/2} \; \Gamma\left( {11 \over 2} \right)} \,
\left[1- {x^2 \over 22} + {x^4 \over 1144}  + {\mathcal{O}}\left(x^6\right) \right] \; 
\eeq
which shows that the above solution is regular at the origin and normalizable.

\noindent
{\bf{(b) Equilateral Case with Curvature Term $\left(\epsilon = 0\right)$}}

Next we include again the effects of the curvature term.
Since now the deficit angle $\delta = \pi/3 $ at each vertex, the curvature
contribution for each triangle is 
$ \kappa \cdot { \pi \over 3 } \cdot 3  = \pi \, \kappa $.
For the icosahedron one has in Eq.~(\ref{eq:kappa})
\beq
\kappa_{icosa} \; = \; 2 \cdot {1 \over 5} \; .
\eeq
Then with the curvature term included for equilateral triangles
one obtains for equilateral triangles $[{\mathcal{O}} (\epsilon^0)]$ 
\bea
\Psi ( A_{tot} ) 
& \simeq &
\exp\left ( - \, 2 \, i \, \sqrt{\tilde{\lambda}} \, A_{tot} \right ) 
\; _1F_1 \left ( 5 - i \, 
{5 \, \pi \, \kappa_{icosa} \over \sqrt{\tilde{\lambda}} \, G^2}
,\; 10,\; 4 \, i \, \sqrt{\tilde{\lambda}} \, A_{tot} \right )
\nonumber \\
& = &
\exp \left ( - \, 2 \, i \, \sqrt{\tilde{\lambda}} \, A_{tot} \right ) \;
_1F_1 \left ( 5- i \, 
{2 \, \pi  \over \sqrt{\tilde{\lambda}} \, G^2},
\; 10,\; 4 \, i \, \sqrt{\tilde{\lambda}} \, A_{tot} \right ) \; ,
\eea
up to an overall wave function normalization constant.
Note that in this case one had to include a factor 
$A_{tot}/4 A_\Delta$, which in the dodecahedron case equals five.

\noindent
{\bf{(c) Large Area in the Strong Coupling Limit $\left(\epsilon \ne 0\right)$}}

In the limit of large areas the two independent solutions reduce to
\beq
\Psi
\; \mathrel{\mathop\sim_{ x \; \rightarrow \; \infty } } \; 
\exp \left (\pm \, i \,x \right )
\; \sim \; \exp\left ( \pm \, 2 \, i \, \sqrt{\tilde{\lambda}}  \, A_{tot}
\right ) 
\eeq
to all orders in the weak field expansion parameter $\epsilon$, as
for the tetrahedron and octahedron case.
Note also that in the strong coupling limit the two independent
wave function solutions again completely factorize as a 
product of single-triangle contributions.

\noindent
{\bf{(d) Small Area in the Strong Coupling Limit $\left(\epsilon \ne 0\right)$}}

In the limit of small area, the solution approaches 
a constant in the equilateral case.
To go beyond the equilateral case, one can write again a general ansatz 
for the wave function, written in terms of geometric invariants
as in Eq.~(\ref{eq:ansatz}).
Then the solution in $\epsilon$ for small $s$.
To zeroth order in $\epsilon$ the solution is
$ \Psi \sim J_n (x) / x^n  $ with $n = 9/2 $.
This gives in Eq.~(\ref{eq:ansatz})
$\gamma_0 =0$, $\gamma_2 = - \, {2 \over 11} \, {\tilde{\lambda}}$ and
$\gamma_4  = {2 \over 143} \,  {\tilde{\lambda}}^2$.
But to linear order [${\mathcal{O}}\left(\epsilon \right)$] 
one finds again that linear terms in $h$ appear which cannot be
expressed in the form of  Eq.~(\ref{eq:ansatz}).
But one also finds that, while these terms are nonzero if one uses the Hamiltonian 
density (the Hamiltonian contribution from just a single triangle),
if one uses the sum of such triangle Hamiltonians then the 
resulting solution is symmetrized, and the corrections to
Eq.~(\ref{eq:ansatz}) are found to be of order 
${\mathcal{O}}\left(\epsilon^2 \right)$.
Then the wave function for small area is of the form
\beq
\Psi \, \simeq \,  
1 \, - \, {\textstyle {2 \over 11} \displaystyle}
{\tilde{\lambda}} \, A_{tot}^2  
\, + \, {\textstyle {2 \over 143} \displaystyle}
\, {\tilde{\lambda}}^2 \, A_{tot}^4 \, + \, \dots  \; ,
\eeq
up to terms of ${\mathcal{O}} ( \epsilon )$.

\vskip 40pt

\subsection{Torus}

\label{sec:torus}

Finally we will consider a regularly triangulated torus, which will
consist here of an infinite lattice built out of triangles, with each triangle
having 12 neighboring triangles.
The torus topology is equivalent to requiring periodic boundary
conditions in the two spatial directions.
Of course, one could consider the same type of lattice but with
some other sort of boundary condition, but we shall not pursue
that aspect here.

Due to the local structure of the lattice Wheeler-DeWitt equation in
Eq.~(\ref{eq:wd_2d}),
it will not be necessary to include in the wave function triangles that
are arbitrarily far apart.
Instead it will be sufficient, in order to determine the overall structure of
the solution, to include only those triangles that
are affected in a nontrivial way by the interaction terms in the
Wheeler-DeWitt equation.
In the present case, this requires the consideration of one given
triangle plus its 12 neighbors, giving a total of 13 triangles.
Here, we will also set as before 
$ x  \, \equiv \, 2 \, \sqrt{\tilde{\lambda}}  \, A_{tot} $.

\noindent
{\bf{(a) Equilateral Case in the Strong Coupling Limit $\left(\epsilon = 0\right)$}}

For this case the relevant equation and its solution are largely in line with
what was obtained for the previous cases.
For equilateral triangles the wave function $\Psi $ has to satisfy
\beq
\Psi'' \, + \, { 13 \over 2 \, x } \, \Psi' \, + \, \Psi \, = \, 0 \; .
\label{eq:tor_eq}
\eeq
The wave function can now be written as
\beq
\Psi(x) 
\; = \; \mathcal{N} \, {J_n (x) \over x^n} 
\eeq
with, here, (due to our specific choice of sublattice)
\beq
n\; = \; {11 \over 4}
\eeq
so that 
\beq
\Psi(x) \; = \; \mathcal{N} \; 
{ J_{11/4} \left( 2 \sqrt{\tilde{\lambda}}  \, A_{tot} \right) \over  
\left(          2 \sqrt{\tilde{\lambda}}  \, A_{tot} \right)^{11/4}} \; .
\eeq
The wave function normalization constant is given in this case by
\beq
\mathcal{N} \;  = \;  4 \, \sqrt
{ 30 \, \Gamma\left ({13 \over 4}\right) \over \Gamma \left( {11 \over 4}\right) } \;
\tilde{\lambda} ^{1/4} \; .
\eeq
For the above wave function an equivalent form is
\beq
\Psi ( A_{tot} ) \; = \; 
\mathcal{N} \; {1 \over 2^{11/4} \; \Gamma\left({15 \over 4} \right)} \;
\exp \left ( - \, 2 \, i \, \sqrt{\tilde{\lambda}} \, A_{tot} \right ) \;
_1 F_1 \left ( {13 \over 4}, {13 \over 2} , 4 \, i \, \sqrt{\tilde{\lambda}} \, A_{tot} \right )
\; .
\eeq
Expanding the above solution for small area one obtains
\beq
\Psi \; = \;
\mathcal{N} \; {1 \over 2^{11/4} \; \Gamma\left({15 \over 4} \right)} \;
\left[\, 1 - {x^2 \over 15} + {x^4 \over 570} + {\mathcal{O}}(x^6)
\,  \right ] \; ,
\eeq
which shows the above solution is indeed regular at the origin.

\noindent
{\bf{(b) Equilateral Case with Curvature Term $\left(\epsilon = 0\right)$}}

In the case of the torus, the curvature term is zero ($\chi=0$),
so there are no changes to the preceding discussion.

\noindent
{\bf{(c) Large Area in the Strong Coupling Limit $\left(\epsilon \ne 0\right)$}}

In the limit of large areas the two independent solutions reduce to
\beq
\Psi 
\; \mathrel{\mathop\sim_{ x \; \rightarrow \; \infty } } \;
\exp \left ( \pm \, i \,x \right )
\; \sim \; \exp \left ( \pm \, i \, {2 \, \sqrt{\tilde{\lambda}}}   \, A_{tot}
\right )  \;
\eeq
to all orders in $\epsilon$.
This is similar to what was found earlier for the other lattices.
In particular, the two independent
solutions again completely factorize as a 
product of single-triangle contributions.

\noindent
{\bf{(d) Small Area in the Strong Coupling Limit $\left(\epsilon \ne 0\right)$}}

In the limit of small area, the regular solution approaches a
constant and the discussion, and solution, is rather similar to the previous cases.
Here one finds
\beq
\Psi \, \simeq \,  
1 \, - \, {\textstyle {4 \over 15} \displaystyle}
{\tilde{\lambda}} \, A_{tot}^2  
\, + \, {\textstyle {8 \over 285} \displaystyle}
\, {\tilde{\lambda}}^2 \, A_{tot}^4 \, + \, \dots  \; ,
\eeq
up to terms of ${\mathcal{O}} ( \epsilon^2 )$.

\vskip 40pt

\subsection{Summary of Results}
\label{sec:summary}

In this section we will summarize the results obtained so far for the
various finite lattices considered (tetrahedron, octahedron,
icosahedron, and regularly triangulated torus).

\noindent
{\bf{(a) Equilateral Case in the Strong Coupling Limit $\left(\epsilon = 0\right)$}}

It is rather remarkable that all of the previous cases (except the
trivial case of a single triangle, which has no curvature)
can be described by one single set of interpolating wave functions, where the
interpolating variable is simply related to the overall lattice size
(specifically, the number of triangles).

Indeed for equilateral triangles and in the absence of curvature,
the wave function $\Psi (x) $ for all previous cases is a solution to 
the following equation
\beq
\Psi'' \, + \, { 2 n + 1 \over x } \, \Psi' \, + \, \Psi \, = \, 0 \; ,
\label{eq:psi_eq}
\eeq
with parameter $n$ given by
\beq
n \; = \; \quarter \, ( N_\Delta \, - \, 2 ) 
\label{eq:n_N2}
\eeq
where $N_\Delta \equiv N_2 $ is the total number of triangles on the lattice.
Thus
\beq
N_\Delta \; = \; 4 ( n + \half )
\eeq
and, consequently,
\bea
n_{tetrahedron} & = & \quarter \, ( 4 - 2 ) \, = \, {1 \over 2}  \; ,
\nonumber \\ 
n_{octahedron} & = & \quarter \, ( 8 - 2 ) \, = \, {3 \over 2}  \; ,
\nonumber \\ 
n_{icosahedron} & = & \quarter \, ( 20 - 2 ) \, = \, {9 \over 2} \; ,
\nonumber \\ 
n_{torus} & = & \quarter \, ( 13 - 2 ) \, = \, {11 \over 4} \; .
\eea
Note that for a single triangle one has $n=\half$ as well, but the
definition of the scaled area is different in that case.

Furthermore the differential equation in
Eq.~(\ref{eq:psi_eq}) describes, in spherical coordinates and with
suitable choice of constants,
the radial wave function for a free quantum particle in $D= 2 n + 2$ 
dimensions.
Indeed recall that in $D$ dimensions the Laplace operator in spherical coordinates
has the form
\beq
\Delta \Psi \; = \; 
{ \partial^2 \Psi \over \partial r^2 } \, + \,
{ D - 1 \over r } \, { \partial \Psi \over \partial r } \, + \,
{ 1 \over r^2 } \, \Delta_{S^{D-1}} \, \Psi
\eeq
where $ \Delta_{S^{D-1}} $ is the Laplace-Beltrami operator
on the $(D-1)$-sphere.
In our case, the wave function does not, to this order, depend on
angles and therefore the last (angular variable) term does not contribute.
The role of the angles is played in our case by the $h$ variables,
which to this order do not fluctuate.

A nonsingular, normalizable solution to Eq.~(\ref{eq:psi_eq})
is then given by
\beq
\Psi (x) \; = \; \mathcal{N} \; { J_{n} (x) \over x^n } 
\; = \; \tilde{\mathcal{N}} \;
e^{- i \, x} \; _1F_1 \left ( n+ \half , 2 \, n+1  , 2 \, i \, x \right )
\label{eq:wf_zeroR}
\eeq
where $ \mathcal{N} $ is the wave function normalization constant
\beq
\mathcal{N} \;\equiv \; 2 \, \left [ 
{ \Gamma ( n + \half ) \, \Gamma ( 2 n + \half ) \over \Gamma (n) } 
\right ]^{1/2} \, 
\tilde{\lambda} ^{1/4} \; ,
\label{eq:norm1}
\eeq
and
\beq
\tilde{ \mathcal{N} } \; \equiv \; { 1 \over 2^n \, \Gamma ( n+1 ) } \;
\mathcal{N} \; .
\label{eq:norm2}
\eeq
Here and in Eq.~(\ref{eq:wf_zeroR}) $_1F_1 (a,b;z)$ denotes the 
confluent hypergeometric function of the first kind,
sometimes denoted also by $M (a,b,;z)$.
In either form, the above wave function is real, in spite of appearances.
The general asymptotic behavior of the solution $\Psi (x)$ is found 
from Eq.~(\ref{eq:psi_eq}).
For small $x$ one has
\beq
\Psi (x) \; \sim \; x^\alpha
\eeq
with index $\alpha = 0, \, - 2 n$. 
The latter solution is singular and will be discarded.
For large $x$ one finds immediately
\beq
\Psi (x) \; \sim \; { 1 \over x^{n+ \half} } \; \exp \left ( \pm i x \right ) ,
\eeq
which is of course consistent with all the previous results.
Indeed the other possible independent solution of Eq.~(\ref{eq:psi_eq}) would be
\beq
\Psi (x)\; \simeq \; { Y_{n} (x) \over x^n } \; ,
\eeq
where $ Y_n (x) $ is a Bessel function of the second kind (or Neumann function).
However, the latter leads to a wave function $\Psi$ which is singular as $ x \rightarrow 0 $, 
\beq
\Psi (x)\; \sim \; - { 1 \over \pi } \, \Gamma (n) \, 
2^n \, x^{-2 n}
\eeq
and gives, therefore, a solution which is not normalizable.
For completeness we record here the small $x$ (small area) 
behavior of the normalized wave function in Eq.~(\ref{eq:wf_zeroR})
\beq
\Psi (x) \; \sim \; \mathcal{N} \; 
{ 1 \over 2^n \, \Gamma ( n+1 ) } ,
\label{eq:smallx}
\eeq
and the corresponding large $x$ (large area) behavior
\beq
\Psi (x) \; \sim \; \mathcal{N} \; 
\sqrt { 2 \over \pi } \, { 1 \over x^{n+\half} } \, 
\cos \left ( x - { n \pi \over 2 } - { \pi \over 4 } \right ) \; ,
\label{eq:largex}
\eeq
both of which reflect well-known properties of the Bessel functions $J_n (x) $.

\noindent
{\bf{(b) Equilateral Case with Curvature Term $\left(\epsilon = 0\right)$}}

When the curvature term is included in the Wheeler-DeWitt equation,
and still in the limit of equilateral triangles,
one obtains the following interpolating differential equation
\beq
\Psi'' \, + \, { 2 n + 1 \over x } \, \Psi' \, - \,
{ 2 \beta \over x } \, \Psi \, + \, \Psi \, = \, 0 \; ,
\label{eq:psi_eq_R}
\eeq
which now describes the radial wave function for a quantum particle 
in $ D = 2 \, n + 2$ dimensions, with a repulsive Coulomb potential 
proportional to $ 2 \beta $.
The nonsingular, normalizable solution is now given by
\beq
\Psi (x) \; \simeq \; e^{- \, i\, x} \; _1F_1 
\left ( n + \half - i \, \beta , \, 2 \, n+1 , \, 2 \, i \, x \right ) \; ,
\label{eq:wavef_R}
\eeq
up to an overall wave function normalization constant $\tilde{\cal N} (n,\beta)$.
The normalization constant can be evaluated analytically but has a
rather unwieldy form and will not be recorded here.
Note that the imaginary part ($\beta$) of the first argument in the
confluent hypergeometric function of Eq.~(\ref{eq:wavef_R}) 
depends on the topology but does not depend on the number of
triangles.
In view of the previous discussion the parameter $n$ 
increases gradually as more triangles are included in the simplicial geometry.
For the regular triangulations of the sphere,
the total deficit angle (the sum of the deficit angles in a given
simplicial geometry) is always $4 \pi$, so even if one writes for
the wave functional $\Psi[A_{tot}, \delta_{tot}]$, the curvature
contribution $\sum_h \delta_{h}$ is a constant and does not contribute
in any significant way.
Note also that, in spite of appearances, the above wave function is
still real for nonzero $\beta$.
That $\Psi (x) $ in Eq.~(\ref{eq:wavef_R})
is a real function can be seen, for example, from its definition via
the power series expansion
\beq
\Psi (x) \; \simeq \; 
1 \, + \, \frac{2 \beta }{2 n+1} \, x \, - \,
\frac{ 1 + 2 n - 4 \beta ^2 }{4 + 12n + 8 n^2 } \, x^2 \, - \, 
\frac{ \beta  \left(5 + 6n -4 \beta^2 \right) }
{6 \left(3 + 11n +12 n^2 + 4 n^3 \right)}\, x^3 \, + \, 
{\cal  O}(x^4) \; ,
\label{eq:wavef_pow}
\eeq
and again up to an overall normalization factor ${\cal N} (n, \beta)$.

The general asymptotic behavior of the solution $\Psi (x)$ is again easily
determined from Eq.~(\ref{eq:psi_eq_R}).
For small $x$ one has
\beq
\Psi (x) \; \sim \; x^\alpha
\eeq
with again $\alpha = 0, \, - 2 n$, and, therefore, independent of the curvature
contribution involving $\beta$.
The second solution is singular and will be discarded as before.
For large $x$ one finds immediately
\beq
\Psi (x) \; \sim \; { 1 \over x^{n+ \half} } \; \exp \left \{ \,
\pm \, i \, ( x - \beta \ln x ) \, \right \} ,
\eeq
which is of course consistent with all previous results.
It also shows that the convergence properties of the wave function
at large $x$ are not affected by the $\beta$ term.
A second independent solution to Eq.~(\ref{eq:psi_eq_R}) is given by 
\beq
\Psi (x) \; \simeq \; e^{- \, i\, x} \; U \, 
\left ( n + \half - i \, \beta , \, 2 \, n+1 , \, 2 \, i \, x \right ) \; ,
\label{eq:wavef_R_sing}
\eeq
where $U(a,b,;z)$ is the confluent hypergeometric function of the
second kind (sometimes referred to as Tricomi's function).
This second solution is singular at the origin, leading to
a wave function that is not normalizable and will not
be considered further here.

The asymptotic behavior of the regular solution for large argument $z$
(discussed in standard quantum mechanics textbooks such as \cite{ll77,sh85}
and whose notation we will follow here) can be obtained from the asymptotic
form of the confluent hypergeometric function $_1 F_1 $, defined 
originally, for small $z$, by the series
\beq
_1 F_1 (a,b,z) \; = \; 
1 + { a z \over b \, 1!} 
+ { a (a+1) z^2 \over  b(b+1) \, 2! } + \cdots \; .
\eeq
It is common procedure to then write 
$_1 F_1 (a,b,z) = W_1 (a,b,z) + W_2(a,b,z) $, where
$W_1$ and $W_2$ are separately solutions of the 
confluent hypergeometric equation
\beq
z { d^2 F \over d z^2 } + (b-z) { d F \over d z } 
- a \, F \, = \, 0 \; .
\eeq
Then an asymptotic expansion for $_1 F_1 $ (or $M$) is obtained from the
following relations:
\bea
W_1 (a,b,z) & = & { \Gamma (b) \over \Gamma (b-a) } \, ( - z )^{-a} 
\; w (a,a-b+1,-z)
\\ \nonumber
W_2 (a,b,z) & = & { \Gamma (b) \over \Gamma (a) } \, e^z \,  z^{a-b} 
\; w (1-a,b-a,z)
\eea
where
\beq
w (\alpha, \beta, z) \; \mathrel{\mathop\sim_{ z \; \rightarrow \; \infty } } \; 
1 + { \alpha \beta \over z \, 1!} 
+ { \alpha ( \alpha + 1) \beta (\beta+1) \over z^2 \, 2! } + \cdots \; ,
\eeq
with the irregular (at the origin) solution given instead by the 
combination $ G (a,b,z) = i \, W_1 (a,b,z) - i \, W_2(a,b,z) $.
One immediate and useful consequence of the above result is that, as
anticipated before, the behavior of
the regular solution close to the origin is not affected
by the presence of the $\beta$ (curvature) term.
In other words, the wave function solution $\Psi (x) $ in
Eq.~(\ref{eq:wavef_R})
is always well behaved for small areas and, therefore, leads
to a perfectly acceptable, normalizable solution.

Furthermore, the combination and properties of arguments in the 
confluent hypergeometric function in 
Eq.~(\ref{eq:wavef_R})
allow one to write it equivalently as a Coulomb wave function with
(Sommerfeld) parameter $\eta$
\beq
C_l ( \eta ) \; \rho^{l + 1} \cdot
e^{- \, i \, \rho} \, _1F_1 \left ( l+1 - \, i \, \eta, \, 2 \, l + 2, \,
2 \, i \, \rho  \right ) \; = \; F_l ( \eta , \rho ) \, ,
\label{eq:coulomb}
\eeq
where $F_l ( \eta, \rho) $ denotes the {\it regular} Coulomb wave function
that arises in the solution of the quantum mechanical
three-dimensional Coulomb problem in spherical coordinates \cite{ll77,sh85}.
The latter is a solution of the radial differential equation
\beq
{ d^2 \, F_l \over d \, \rho^2 } \, + \, \left [ 1 \, - \, { 2 \, \eta \over \rho } \, - \, 
{ l ( l+1) \over \rho^2 } \right ] \, F_l \; = \; 0 \; ,
\label{eq:radial_eq}
\eeq
with the actual radial wave function then given by $ R_l (r) = F_l ( k r) / r $. 
After comparing the above equation with Eq.~(\ref{eq:wavef_R}) one 
then identifies $ \rho = x $, $l = n - \half $ and $\eta = \beta$.
Thus $l = N_\Delta / 4 - 1 $, where $ N_\Delta $ is the number of
triangles on the lattice.
The proportionality constant $C_l$ in Eq.~(\ref{eq:coulomb}) is given
by the (Gamow) parameter
\beq
C_l ( \eta ) \; \equiv \;
{2^l \, e^{- {\pi \, \eta \over 2}} \, 
\vert \Gamma ( l+1 + \, i \, \eta ) \vert \over \Gamma ( 2l + 2 ) } \;.
\label{eq:coulomb1}
\eeq
One then has immediately, from Eq.~(\ref{eq:wavef_R}), an equivalent
representation for the regular wave function as
\beq
\Psi (x) \; \simeq \;  \left [ C_{ n - \half } ( \beta ) \right ]^{-1} \, 
{ 1 \over  x^{n+\half} } \, F_l ( \beta , x ) \; ,
\label{eq:wavef_coul}
\eeq
again up to an overall wave function normalization constant $\tilde{\cal N} (n,\beta)$.
Again we note here that, on the other hand, the {\it irregular} Coulomb wave function 
[usually denoted by $G_l (\eta, \rho )$]
is singular for small $r$ and will, therefore, not be considered here.
Further relevant properties of the Coulomb wave function can be found
in \cite{ll77,sh85,as72,as12,nist10}.

The known asymptotics of Coulomb wave function \cite{as72,as12,nist10}
allow one to derive the following result for the wave 
function $\Psi $ for large $x$
\beq
\Psi (x) \; \simeq \; \tilde{\cal{N}} \; 
{ 1 \over C_{ n- \half} ( \beta ) \cdot x^{n + \half } } \;
\sin \left [ x - \beta \ln 2 x - { ( 2 n- 1 ) \pi \over 4 } + \sigma_n
\right ]
\label{eq:largex_R}
\eeq
with (Coulomb) phase shift
\beq
\sigma_n (\beta) \; = \; \arg \Gamma ( n + \half + i \beta ) \; .
\eeq
Also, from Eq.~(\ref{eq:coulomb1}),
\beq
C_{n-\half} ( \beta ) \; \equiv \;
{2^{n-\half} \, e^{- {\pi \, \beta \over 2}} \, 
\vert \Gamma ( n + \half + \, i \, \beta ) \vert \over \Gamma ( 2n + 1 ) } \;.
\eeq
It is easy to check that the above result correctly reduces to the
asymptotic expression given earlier for $\Psi$ in Eq.~(\ref{eq:largex}) 
in the limit $\beta=0$. 
The structure of the wave function in Eq.~(\ref{eq:largex_R}) implies 
that the norm is still finite for $\beta \neq 0$,
since the convergence properties of the wave function are not
affected by the curvature term.

\noindent
{\bf{(c) Large Area in the Strong Coupling Limit $\left(\epsilon \ne 0\right)$}}

In the limit of large areas the two independent solutions reduce to
\beq
\Psi
\; \mathrel{\mathop\sim_{ x \; \rightarrow \; \infty } } \;
\exp\left ( \pm \, i \,x \right )
\label{eq:wf_large}
\eeq
where $ x \propto A_{tot} $.
This is true without assuming the weak field expansion, as
was already the case before (see, in particular, the section discussing
the tetrahedron case).

Consequently in the strong coupling limit the two wave function solutions in
Eq.~(\ref{eq:wf_large}) completely factorize as a product of
single-triangle contributions,
\beq
\Psi \; \simeq \; \prod_\Delta \, 
\exp \left ( \pm \, 2 \, i \, \sqrt{\tilde{\lambda}} \, A_{\Delta} \right )  \; ,
\label{eq:wf_large1}
\eeq
again up to an overall normalization constant.
The above result, anticipated in \cite{hw11}, was the basis for the
variational treatment using correlated product wave functions
given in our previous work.
Note also, in view of the result of Eq.~(\ref{eq:largex}), that the
correct solution, satisfying the required regularity condition for small areas,
is actually a linear combination of the above factorized solutions.

\noindent
{\bf{(d) Small Area in the Strong Coupling Limit $\left(\epsilon \ne 0\right)$}}

In the limit of small area, we have shown before in all cases that
the solution reduces to a constant in the equilateral case 
[$ {\mathcal{O}}\left(\epsilon^0 \right) $] for
small $x$ or small areas.
To linear order [${\mathcal{O}}\left(\epsilon \right)$]
the general result is still that linear terms in $h$ appear which cannot be
expressed in the form of  Eq.~(\ref{eq:ansatz}).
But one also finds that, while these terms are nonzero if one uses the Hamiltonian 
density (the Hamiltonian contribution from just a single triangle),
if one uses the sum of such triangle Hamiltonians then the 
resulting solution is symmetrized, and the corrections to
Eq.~(\ref{eq:ansatz}) are found to be of order 
${\mathcal{O}}\left(\epsilon^2 \right)$.
In other words, it seems that some residual lattice artifacts that survive at very
short distances can be partially removed by a suitable coarse-graining procedure
on the Hamiltonian density.


One might wonder what lattices correspond to values of $n$ greater
that 9/2, which is the highest value attained for a regular
triangulation of the sphere, corresponding to the icosahedron.
For each of the three regular triangulations with $N_0$ sites 
one has for the number of edges $N_1 = { q \over 2 } N_0 $ and for the
number of triangles $N_2 = ( { q \over 2 } -1 ) N_0 + 2 $, where
$q$ is the number of edges meeting at a vertex (the local coordination
number).
In the three cases examined before $q$ was
between three and five, with six corresponding to the regularly triangulated torus.
Note that for a sphere $N_0 -N_1 +N_2 = 2$ always.
The interpretation of other, even noninteger, values of $q$ is then clear.
Additional triangulations of the sphere can be constructed by
considering
irregular triangulations, where now the parameter $q$ is interpreted as
an {\it average} coordination number.  
Of course the simplest example is a semiregular lattice with $N_a$ vertices with
coordination number $q_a$ and $N_b$ vertices with coordination
number $q_b$, such that $N_a+N_b = N_0 $.
Various irregular and random lattices were considered 
in detail some time ago in \cite{itz83},
and we refer the reader to this work for a clear exposition
of the properties of these lattices.

We conclude this section by briefly summarizing the key properties of the
gravitational wave function given in Eqs.~(\ref{eq:wavef_R}) and
(\ref{eq:wavef_coul}), 
which from now on will be used as the basis for additional calculations.
First we note that the above wave function is a function of the total
area and total curvature only and, as such, is manifestly 
diffeomorphism-invariant and in accord with the spatial 
diffeomorphism constraint.
While it was derived by looking at the discrete triangulations of
the sphere, it contains a parameter $n$, related to the total number of
triangles on the lattice by Eq.~(\ref{eq:n_N2}), that will allow us
to go beyond the case of a finite lattice and investigate the
physically meaningful, and presumably universal, infinite volume
limit $n \rightarrow \infty$ [see Eq.~(\ref{eq:infvol})].
We have also shown that the above wave function is, in all cases,
an exact solution of the full 
lattice Wheeler-DeWitt equation of Eq.~(\ref{eq:wd_latt1}) in the limit 
of large areas, and to all orders in the weak field expansion. 
Again, this last case is most relevant for taking the infinite volume limit,
defined previously in Eq.~(\ref{eq:infvol}).
Furthermore, the small area behavior of the wave function
plays a crucial role in uniquely constraining, through the regularity
condition, the correct choice of solution.
In this last limit one also finds that the various individual 
lattice solutions agree with the universal form of Eqs.~(\ref{eq:wavef_R}) and
(\ref{eq:wavef_coul}) only to a low order in the weak field
expansion, which is expected given the different short distance
lattice artifacts of the regular triangulation solutions.
Nevertheless, knowledge of their behavior is completely adequate
for extracting the most important physically relevant piece of information, namely 
the constraint on the wave function 
based on the stated regularity condition at small areas, which
comes down to a simple integrability or power counting argument.

\vskip 40pt

\section{Average Area}
\label{sec:avearea}

\begin{figure}
\begin{center}
\includegraphics[width=.7\textwidth]{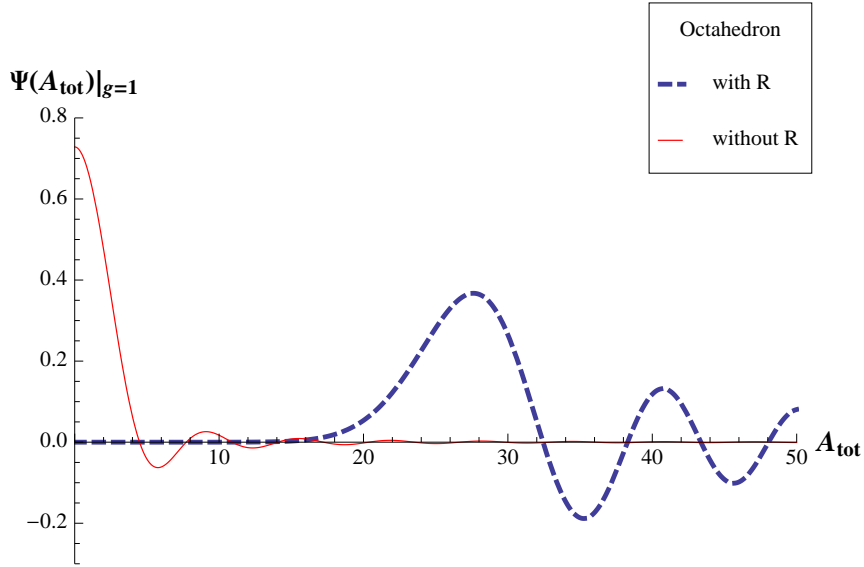}
\end{center}
\caption{\label{fig:octa_psi}Wave Function $\Psi$ versus total
area for the octahedron lattice, with and without curvature contribution.
The wave function is shown here for $g = \sqrt{G} = 1$,
a value chosen here for illustration purposes.
The relevant expression for the wave function is given in
Eq.~(\ref{eq:wavef_R1}). 
We refer to the text for further details on how the wave function
was obtained, and what its domain of validity is.
The wave functions shown here have been properly normalized.
Note that with a nonzero curvature term the peak in the wave function
moves away from the origin.}
\end{figure}

\begin{figure}
\begin{center}
\includegraphics[width=.7\textwidth]{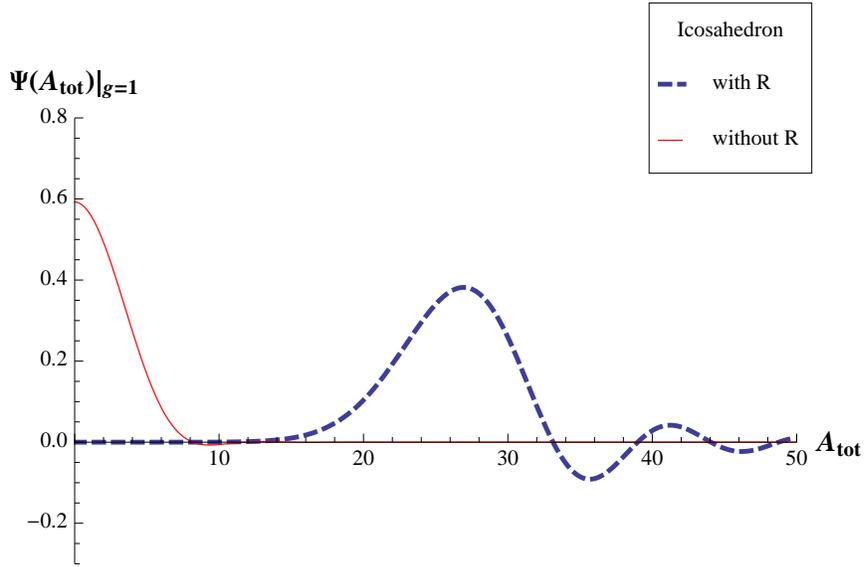}
\end{center}
\caption{\label{fig:icos_psi}Same wave function $\Psi$ as in
Fig.~\ref{fig:octa_psi}, but now for the icosahedron lattice.}
\end{figure}

In this section we will look at a natural quantum mechanical
expectation value, the average total
physical area of the lattice simplicial geometry.
It is one of many quantities
that can be calculated within the lattice quantum gravity formalism, and
is clearly both manifestly geometric and diffeomorphism-invariant.
Here we will use the wave functions given in
Eqs.~(\ref{eq:wavef_R}) and (\ref{eq:wavef_coul}), originally obtained for the 
tetrahedron, octahedron and icosahedron, and later extended
to any number of triangles $N_\Delta$
\beq
\Psi ( A_{tot} ) \; \simeq \; e^{- \, i\, {A_{tot} \over g} } \, 
_1F_1 \left( n + \half - i \, \beta \,, 2 \, n+1 , 2 \, i \, { A_{tot} \over g} \right)\; ,
\label{eq:wavef_R1}
\eeq
with $n \equiv \quarter \, ( N_\Delta \, - \, 2 )$, 
$ \beta \equiv 4 \, \pi  / g^3 $ and $g \equiv \sqrt{G} $, and again valid up to an
overall wave function normalization constant.
Due to the structure of the wave function the resulting probability
distribution for the area is rather nontrivial, having many peaks
associated with the infinitely many minima and maxima of the hypergeometric function.
Clearly the most interesting limit is one where one considers an
infinite number of triangles, $N_\Delta \rightarrow \infty $, which
corresponds to $ n \rightarrow \infty $ in Eq.~(\ref{eq:wavef_R1}).
In Figs.~\ref{fig:octa_psi} and \ref{fig:icos_psi},
we display the behavior of the wave function in
Eq.~(\ref{eq:wavef_R1}), both with and without the
curvature contribution in the Wheeler-DeWitt equation.
One notices that when the curvature term is included $(\beta \neq 0)$, 
the peak in the wave function shifts away from the origin.
This is largely expected, based on the contribution from the 
repulsive Coulomb term in the wave equation of Eq.~(\ref{eq:psi_eq_R}).

The average total area can then be computed from the above wave
function, as the ground state expectation value
\beq
< A >  \; = \;
{\langle \Psi \vert \, A \, \vert \Psi \rangle \over 
\langle \Psi \vert \Psi  \rangle} 
\, = \; { \int \, d \mu [g]  \; A ( g) \; 
\vert \Psi (g) \vert^2 
\over \int \, d \mu [g] \; \vert \Psi (g) \vert^2  } \; ,
\label{eq:ave_area}
\eeq
where $g$, here, is the three-metric, and $ d \mu [g] $ denotes a functional integration
over all three-metrics.
In our case we use the measure
\beq
\int \, d \mu [g]  \;\; \longrightarrow  \;\; \int_0^\infty \, d A_{tot}
\; ,
\label{eq:measure}
\eeq
which then gives, in terms of the scaled area variable $x$,
\beq
< A_{tot} >  \; = \; g \; 
{ \int_0^\infty \, dx \, x \cdot \vert \Psi (x) \vert^2 
\over \int_0^{\infty} \, dx \, \vert \Psi (x) \vert^2  } \;.
\label{eq:ave_area_x}
\eeq
In the absence of a curvature term in the Wheeler-DeWitt equation
$(\beta=0)$, the average area can easily be computed analytically in terms 
of Bessel function integrals, and the result is
\beq
< A_{tot} > \; = \; g \cdot 
\frac{\pi \,(4 n-1) \, \Gamma (4 n-2)} {2^{8 n-5}  \, \Gamma (n)^4} \; .
\eeq
Note that the average area diverges as $n \rightarrow \half$, which corresponds
to the tetrahedron;
this entirely spurious divergence prevents us from using
the tetrahedron lattice in plotting
and numerically extrapolating the remaining two lattices (octahedron and
icosahedron) to the infinite lattice limit.
For the octahedron one finds $<A_{tot}> = 15 \, g / \pi $,
for the icosahedron $<A_{tot}>= 21879 \, g / 3920 \pi $,
and in the large $n$ limit 
$<A_{tot}>= \sqrt{2 n / \pi } \; g + {\cal O} (1 / \sqrt{n})$.

One finds that in the presence of a curvature term ($\beta \neq 0$) the resulting
integrals are significantly more complicated.
We have, therefore, resorted to a number of tools,
which include an analytic expansion in $\beta$, the use of
known asymptotic expansions for the wave function at large arguments,
and an exact numerical integration of the resulting integrals.
Let us first discuss here the expansion in $\beta$.
It is known that the Coulomb wave functions can be expanded in
terms of spherical Bessel functions (Neumann expansion)
\cite{as72,as12,nist10}, so that one has
\beq
F_l (\eta, \rho) \; = \; 
{ 2^{l+1} \over \sqrt{\pi} } \; \Gamma ( l + \thrha ) \;
C_l (\eta) \; \rho \; \sqrt{ \pi \over 2 \rho } \,
\cdot \left \{ \, \sum_{k=l}^\infty \, b_k (\eta ) \, 
J_{k+\half} (\rho ) \right \}
\label{eq:bessel_exp}
\eeq
with coefficients $b_k (\eta) $ given by a simple recursion
relation.
When written out explicitly, the expression in curly brackets
involves
\beq
J_{l+\half} (x) \, + \,  
{2 l + 3 \over l + 1} \, \eta \cdot J_{l+\thrha} (x) \, + \,
{2 l + 5 \over l + 1} \, \eta^2 \cdot J_{l+\fivha} (x) \, + \, \cdots \; ,
\eeq
with additional terms linear in $\eta$ reappearing at higher orders.
That the above expansion is a bit problematic is not entirely surprising,
given the modified asymptotic behavior of the Coulomb
wave functions for $\eta \neq 0$.
In the following, in order to provide initially some insight into 
the effects of the $\eta$ (or $\beta$) term on the wave function 
$\Psi$, we will include the first correction as a perturbation, and
drop the rest.
Later on, higher order corrections can be included as additional
contributions.
With this truncation, the Coulomb wave function in Eq.~(\ref{eq:coulomb}) becomes
\beq
F_l ( \eta, \rho ) \; = \; { 2^{l+1} \over \sqrt{\pi} } 
\, \Gamma \left ( l+ \thrha \right ) \, 
C_l (\eta) \; \rho \; \sqrt{ {\pi \over 2 \, \rho }} \,  
\left [ \, J_{l + \half } ( \rho ) \, + \, \eta \, { 2 l + 3 \over l+1 } 
\, J_{l + \thrha } ( \rho ) \, + \, \cdots \right ]
\label{eq:wavef_trunc0}
\eeq
with the last term treated as a perturbation,
giving for the wave function itself [see Eq.~(\ref{eq:wavef_R})]
\bea
\Psi (x) & \simeq & e^{- \, i\, x} \; _1F_1
\left ( n + \half - i \, \beta , \, 2 \, n+1 , \, 2 \, i \, x \right )
\nonumber \\
 & = & { 1 \over x^n } \, \left [
J_{n} ( x ) \, + \, \beta \, { 2 n + 2 \over n + \half } \,
J_{n + 1 } ( x ) \, + \, \cdots \right ] \; ,
\label{eq:wavef_trunc}
\eea
again up to an overall wave function normalization constant
$\tilde{\cal N} $.
Note that if $m$ Bessel function terms are kept in Eq.~(\ref{eq:wavef_trunc}),
beyond the zeroth order, strong coupling, term involving $J_n(x)$,
then the resulting expansion in $\beta$ contains terms up to $\beta^m$.
One finds to lowest order ($m=1$)
\beq
{ 1 \over \tilde{\mathcal{N}}^2 } \; = \; 
\frac{\Gamma (n)}{2 \Gamma 
\left(n+\frac{1}{2}\right) \Gamma \left(2  n+\frac{1}{2}\right)}
\, + \, \frac{4^{1-n} (n+1) \beta }{(2 n+1) \Gamma (n+1)^2}
\, + \, \cdots
\eeq
From the above expressions, the average area can then be computed 
as some still rather complicated function,
\beq
< A_{tot} >  =  g \, \left \{
 \frac{\pi \,(4 n-1) \, \Gamma (4 n-2)} {2^{8 n-5}  \, \Gamma (n)^4}
 + \frac{4 (n+1) \beta}{2 n+1} \,
\left [ 1-\frac{4^{1-2 n} \Gamma
   \left(n-\frac{1}{2}\right) \Gamma \left(n+\frac{1}{2}\right)
   \Gamma \left(2 n+\frac{1}{2}\right)^2}{n^2 \Gamma
   (n)^6}\right ] + \cdots
\right \} \; .
\label{eq:avar_1}
\eeq
Additional terms can later be included in the Bessel function
expansion of Eq.~(\ref{eq:bessel_exp}), so as to obtain more accurate
values for the averages; this will be done later.

\begin{figure}
\begin{center}
\includegraphics[width=.7\textwidth]{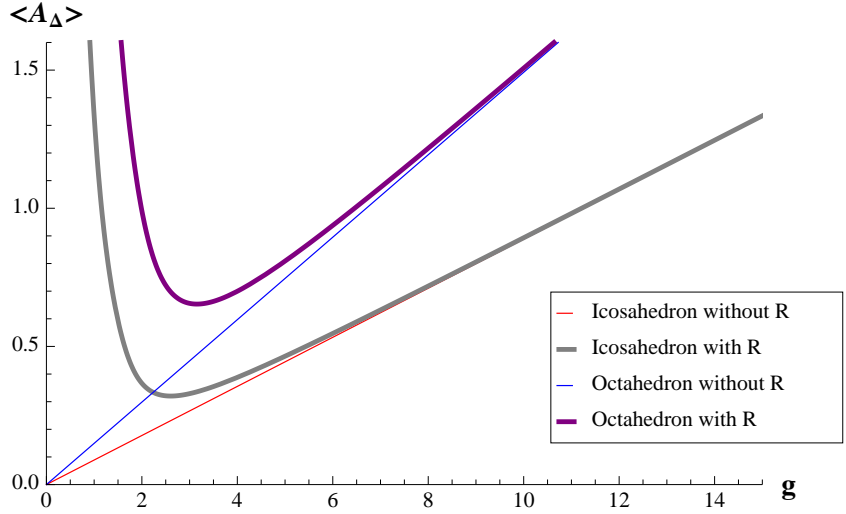}
\end{center}
\caption{\label{fig:ave_area}
Average area of a single triangle vs. 
$g = \sqrt{G}$
for the octahedron and the icosahedron configurations.
The average area was calculated using the expression in
Eq.~(\ref{eq:ave_area_x}).
Note the qualitative change when one includes the curvature term,
with a minimum appearing at $ g \sim {\mathcal{O}} (1) $. 
}
\end{figure}

Figure \ref{fig:ave_area} shows the exact value of the
average area for a single triangle 
$<A_\Delta > = <A_{tot}> / N_\Delta $ as a function of the coupling $g$, 
obtained by doing the integral in Eq.~(\ref{eq:ave_area_x})
numerically, with the wave function given in Eq.~(\ref{eq:wavef_R1}).
One noteworthy aspect is that a qualitative change seems to occur
when one includes the curvature term:
a well defined minimum occurs at 
$ g \sim 1 $, which would suggest the appearance of 
some sort of phase transition.
Doing the integrals numerically one finds
a minimum in the average area of a triangle at $g_c \approx 3.1 $ 
for the octahedron, and at $g_c \approx 2.6 $ for the icosahedron.
On the other hand, using the lowest order Bessel function expansion 
of Eq.~(\ref{eq:wavef_trunc}) for the octahedron ($n=3/2$) one
finds a minimum at $g_c = 2.683$, and for the icosahedron
($n=9/2$) at $g_c = 2.271$.
Adding one more Bessel function correction term then gives
$g_c = 3.135$ and $g_c = 2.637$ for the two cases,
respectively, which suggests that the expansion is converging.

The limit of a large number of triangles 
$N_\Delta \rightarrow \infty $ corresponds to taking the
parameter $n$ in Eq.~(\ref{eq:wavef_R1}) to infinity,
since $ n \equiv \quarter \, ( N_\Delta \, - \, 2 )$.
From the lowest order Bessel function expansion one 
obtains the following analytic expression for the
average total area
\beq
< A_{tot} > \;\, = \; g \cdot \sqrt{ 2 \, n \over \pi } \,
\left [ 1 \, + \, { 3 \over 16 \, n } \, + \, 
{\cal O} \left ( { 1 \over n^2 } \right ) 
\right ] \, + \,
{ 2 ( \pi - 2 ) \, g \over \pi } \, \beta \, + \, \cdots \; ,
\label{eq:avar_2}
\eeq
with $ \beta \equiv 4 \pi / g^3 $ [see Eq.~(\ref{eq:beta_g})].
In this limit the resulting function of $g$ has, again, a well defined minimum at 
\beq 
g_c^3 \; = \; { 8 ( \pi - 2 ) \, \sqrt{2 \pi} \over \sqrt{n} } \; 
\label{eq:gc}
\eeq
or $g_c \simeq 2.839 / n^{1/6}$ for large $n$ with one Bessel
function correction term.
With two Bessel function correction terms in Eq.~(\ref{eq:wavef_trunc})
one finds $g_c \simeq 3.276 / n^{1/6}$,
which again suggests that the expansion is slowly converging.
Using the exact wave function to do the integrals numerically one
finds for the minimum $g_c \simeq 3.309 / n^{1/6} $, which is close to the above answer.
Interestingly enough, the above result would suggest that in the limit of infinitely
many lattice points the critical point $g_c$ actually moves to the origin,
indicating a phase transition located at exactly $g=0$ ($G=0$)
in the infinite volume ($ n \rightarrow \infty $) limit 
(see further discussion later).
We note here that the average area for a single triangle is obtained by simply
dividing the average total area by the total number of
triangles $N_\Delta = 4 n + 2 $, which then gives in the same limit
of large $n$ and strong coupling
\beq
< A_{\Delta} > \; = \; { g \over 2 \, \sqrt{ 2 \, \pi \, n } }
\, + \, {\cal O} ( { 1 \over n } ) \; .
\label{eq:avar_3}
\eeq
Quite generally, the average of the area per site in the lattice theory
(the spatial volume per site) appears to be well defined  
mainly due to our wave function normalization choices and, consequently,
can be explicitly calculated without any leftover ambiguity.

As will be discussed further below in more detail, the estimate for the critical
point given in Eq.~(\ref{eq:gc}) is also in good agreement with
a previous variational estimate.
In \cite{hw11} the quantum-mechanical variational (Rayleigh-Ritz) method was used to find an 
approximation for the ground state wave function, using
as variational wave function a correlated (Jastrow-Slater) product of 
single-triangle wave functions.
There it was found, from the roots of the equation 
$ < \Psi \vert H \vert \Psi > \, = 0$, 
that the variational parameters are almost purely imaginary for 
strong coupling (large $G > G_c$), 
whereas for weak enough coupling (small $G < G_c$) they become real.
This abrupt change in behavior of the wave function at
$G_c$ then suggested the presence of a phase transition.
With the notation used in this paper, the result of \cite{hw11} reads
$ g_c^3 \sim 1 / N_\Delta $, in qualitative agreement with
the result of Eq.~(\ref{eq:gc}), in the sense that both
calculations point to a critical point $G_c = 0$ in the infinite volume limit.

Let us now make some additional comments which should help clarify
the interpretation of the previous results.
It is well known that if there is some sort of continuous phase transition in the
lattice theory, the latter is generally associated 
with a divergent correlation length in the vicinity of the critical
point.
In our case it is clear that at strong coupling (large $g$)
the correlation length is small (of order one)
in units of lattice spacing.
This can be seen from the fact that (a) the coupling term in the
Wheeler-DeWitt equation is due mainly to the curvature term, which is 
small for large $g$, and (b) that the ground state wave function
is of the form of a correlated product in the same
limit [see Eq.~(\ref{eq:wf_large1})].
Then as the effects of the curvature term are included, the correlation length
starts to grow due to the additional coupling between edge variables.
The previous calculation would then suggest that the point
of divergence is located at $g=0$.
It is, of course, essential that one looks at the limit of
infinitely many triangles, $N_\Delta \rightarrow \infty $, 
since no continuous phase transition can occur in a system
with a finite number of degrees of freedom.

It is also of interest here to discuss how the above (Lorentzian) results relate
to what is known about the corresponding {\it Euclidean} lattice
theory in three dimensions, which was studied in some detail in \cite{hw93}.
There a phase transition was found between two phases, with the
weak coupling phase $G<G_c$ exhibiting a sort of pathological behavior,
whereby the lattice collapses into what geometrically could be
described as a branched polymer.
This is clearly a nonperturbative phenomenon that cannot be seen
from perturbation theory in $G$.
In the Euclidean formulation, average volumes are obtained as suitable 
derivatives of $ \log Z_{latt}$ with respect to the bare cosmological constant
$\lambda_0$, where $Z_{latt}$ is the lattice path integral 
\beq
Z_{latt} = \int [dl^2] \, e^{-I_{latt}(l^2)} 
\label{eq:zlatt}
\eeq
with, in four dimensions, the action given by
\beq
I_{latt} = \lambda_0 \, \sum_h V_h(l^2) - k \sum_h \delta_h (l^2) A_h (l^2)
\label{eq:ilatt}
\eeq
and $h$ denoting a hinge [more details can be found in \cite{hw93}].
Similarly, the average curvature can also be obtained as a
derivative of $\log Z$ with respect to  $k \equiv 1/(8 \pi G)$.
More importantly, a nonanalyticity in $Z$, 
as induced by a phase transition, is expected to show up in local averages as well.
From the above expression for $Z_{latt}$, exact sum rules can be
derived relating various averages \cite{ham00}.
In the case of the three-dimensional Euclidean theory the sum rule reads
\beq
2 \, \lambda_0 < \sum_T V_T> - k <\sum_h \delta_h l_h> \, - \, C_0 \; = \;0
\label{eq:sumrule}
\eeq
where the first term contains a sum over all lattice tetrahedra, 
and the second term involves a sum over all lattice hinges (just edges in this case). 
The quantity $C_0$, here, is a constant that solely depends on how the
lattice is put together (i.e. on the local coordination number, or
incidence matrix).

In \cite{hw93} it was found that the average curvature goes to zero
at some $g_c$ with a characteristic universal exponent $\delta$, 
\beq
<\sum_h \delta_h l_h> \; = \; - R_0 \, \left \vert g - g_c \right \vert^{\delta} 
\label{eq:delta_exp}
\eeq
and that the curvature fluctuation diverges in the same limit.
From the sum rule in Eq.~(\ref{eq:sumrule}) one then deduces that the 
average volume in the Euclidean theory has a singularity of the type 
\beq
< \sum_T V_T> \; = \; V_0 \, - \, V_1 \, \left \vert g - g_c \right \vert^{\delta} 
\label{eq:delta_exp1}
\eeq
with the same exponent $\delta \simeq 0.77$.
The latter is related by standard universality and
scaling arguments \cite{par81,car97,zin02} (see \cite{hbook} for details
specific to the gravity case)
to the correlation length exponent $\nu$ by 
$\nu = ( 1 + \delta ) / d $ in  $d$ dimensions.
To compare to the Lorentzian theory discussed in this paper,
one notes that the three-dimensional Euclidean theory corresponds
to the $(2 + 1)$-dimensional Wheeler-DeWitt theory, so
that the average volume in the above discussion should be taken
to correspond to an average area in our case.
\footnote{
It should be noted that in the case of the lattice
Wheeler-DeWitt equation of Eqs.~(\ref{eq:wd_latt}) and
(\ref{eq:wd_latt1}) and, generally, in any lattice Hamiltonian 
continuous-time formulation,
the lattice continuum limit along the time direction has already
been taken.
This is due to the fact that one can view the resulting $2+1$ 
theory as originating from one where there exist initially
two lattices spacings, $a_t$ and $a$.
The first one is relevant for the time direction and the second
one for the spatial directions.
In the present lattice formulation the limit $a_t \rightarrow 0$
has already been taken; the only limit left is $a \rightarrow 0$,
which requires the existence of an ultraviolet fixed point of
the renormalization group.
}
To conclude, the results for the average area suggest the existence
of a phase transition in the Lorentzian theory located at $g=0$.
In the next sections we will present a further test of this hypothesis,
based on physical observables that can establish directly and unambiguously
the location of the phase transition point.

\vskip 40pt

\section{Area Fluctuation, Fixed Point and Critical Exponent}
\label{sec:fluct}

Another quantity that can be obtained readily from the wave function 
$\Psi$ is the fluctuation in the total area
\beq
\chi_A \; = \; { 1 \over N_{\Delta} } \,
\left \{ < ( A_{tot} )^2 > \, - \, < A_{tot} >^2 \right \} \; .
\label{eq:chi_a}
\eeq
The latter is related to the fluctuations in the individual triangles
by
\beq
\chi_A \; = \; N_{\Delta} \, 
\left \{ <  A_{\Delta}^2 > \, - \, < A_{\Delta} >^2 \right \}
\label{eq:chi_a1}
\eeq
with the usual definition of averages, such as the
one given in Eq.~(\ref{eq:ave_area}).

Generally for a field $\phi (x) $ with renormalized mass $m$ and
correlation length $\xi = m^{-1}$, wave function renormalization
constant $Z$, and (Euclidean) propagator
\beq
< \phi (x) \phi (0) > \; = \; \int { d^d p \over ( 2 \pi )^d } \,
e^{-i p \cdot x } \; { Z \over p^2 + m^2 }  \; ,
\eeq
one has for $\Phi \equiv \int_x \phi(x) $
\beq
< \Phi^2 > \; = \; \int_{x,y} \, < \phi (x) \phi (y) > \; = \;  
V \, \int_{x} \, < \phi (x) \phi (0) > \; = \; V \, { Z \over m^2 }
\; = \; V \; Z \, \xi^2  \; .
\eeq
Thus the field fluctuation probes the propagator at zero momentum,
which in turn is directly related to the renormalized mass 
(and thus $\xi$) for the field in question.
If the field $\Phi$ acquires a nonzero expectation value, the above
result is modified to
\beq
{ 1 \over V } \, \left \{ < \Phi^2 > \, - \, < \Phi >^2 \right \} \; = \;
{ Z \over m^2 } \; = \; Z \; \xi^2 \; ,
\eeq
involving instead the connected propagator.
In the gravity case the quantity $A_{tot}$ plays the role of $\Phi$;
if the fluctuation diverges ($\xi \rightarrow \infty$) then one
has a phase transition or an ultraviolet fixed point in quantum field
theory language \cite{hw84,ham00}.

\begin{figure}
\begin{center}
\includegraphics[width=.7\textwidth]{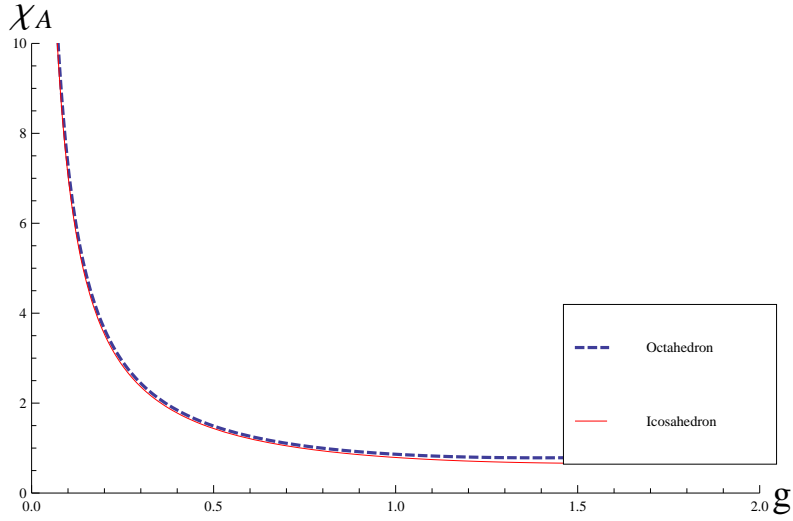}
\end{center}
\caption{\label{fig:chi_a}
Area fluctuation $\chi_A $ vs $g = \sqrt{G}$
for the octahedron and icosahedron, computed from 
Eq.~(\ref{eq:chi_a}).
Note the divergence for small $g$. 
}
\end{figure}

Without the curvature term in the Wheeler-DeWitt equation
[$\beta=0$ for the wave function $\Psi$ in Eq.~(\ref{eq:wavef_trunc})],
the area fluctuation does not diverge, even when $n$ is large
and is simply proportional to $g^2$.
In this case one finds
\beq
\chi_A ( \beta =0 ) \; = \; 
\frac{4 n - 1}{16} \, 
\left [ \frac{2 n-1}{2 n^2 - n - 1}-
\frac{\pi^2 (4 n-1) \Gamma (4 n-2)^2}
{2^{16 n -13} (2 n+1) \Gamma (n)^8} \right ] \, g^2 
\; \sim \; { \pi \, - \, 2 \over  4 \pi } \, g^2 
\, + \, {\cal O} ( { 1 \over n } ) \; .
\eeq
Note the spurious singularity for the special case of the tetrahedron, $n=1/2$.
When the curvature term is taken into account
one finds, from the full wave function $\Psi$ in Eq.~(\ref{eq:wavef_trunc})
and in the limit of large $n$,
\beq
\chi_A \; = \; \left ( 1 - { 2 \over \pi } \right ) \, { g^2 \over 4 }
\, + \, 2 \, ( 4 - \pi ) \sqrt{2 \over n \, \pi } \; { 1 \over g } 
\, + \, \cdots 
\eeq
Note that the fluctuation now appears to diverge as $g \rightarrow 0$
(see also Fig. \ref{fig:chi_a}).
Furthermore, $\chi_A $ is nonanalytic in the original Newton's coupling $ G = g^2 $,
which suggests that perturbation theory in $G$ is useless.
A divergence of the fluctuations as $g \rightarrow 0 $ implies
that in this limit the correlation length diverges 
in lattice units, signaling the emergence of a massless excitation.

Just as for the case of the average curvature [Eq.~(\ref{eq:sumrule})],
an exact sum rule can be
derived in the (Euclidean) lattice path integral formulation, relating
the local volume fluctuations to the local curvature fluctuations.
In the three-dimensional Euclidean path integral theory the following exact identity 
holds for the fluctuations \cite{ham00}
\beq
4 \lambda_0^2 \left [ < (\sum_h V_h)^2 > - < \sum_h V_h >^2 \right ] \; - \; 
k^2 \left [ < (\sum_h \delta_h l_h)^2 > -  < \sum_h \delta_h l_h >^2 \right ]
\; - \; 2 N_1 \; = \; 0 \;\; ,
\label{eq:sumrule1} 
\eeq
where $N_1$ is the number of edges on the lattice 
(further exact sum rules can be derived by considering even higher derivatives
of the free energy $\ln Z_L$ with respect to the parameters $\lambda_0$ and $k$).
Since the last equation relates the fluctuation in the curvature to fluctuations
in the volumes, it also implies a relationship between their singular
(divergent) parts.
\footnote{
We noted previously that in our Hamiltonian formulation,
the lattice continuum limit along the time direction has already
been taken. This results in two lattice spacings, one for the time and one
for the space directions, denoted here respectively by $a_t$ and $a$,
with the first lattice spacing already sent to zero.
It is then relatively straightforward to relate volumes 
between the two formulations, such as $ V \simeq a_t A $. 
Relating curvatures (for example, ${}^2 R $ in the $2+1$ theory
vs the Ricci scalar $R$ in the original three-dimensional
theory) in the two formulations is obviously less easy, due to
the presence of derivatives along the time direction.
}

According to the sum rule of Eq.~(\ref{eq:sumrule1})
a divergence in the curvature fluctuation 
\beq
\chi_R \; \sim  \; < (\sum_h \delta_h l_h)^2 > -  < \sum_h \delta_h l_h >^2 
\eeq
for the three-dimensional (Euclidean) theory
generally implies a corresponding divergence in the volume fluctuation
\beq
\chi_V \; \sim  \; < (\sum_h V_h)^2 > - < \sum_h V_h >^2 \;
\eeq
for the same theory.
In our case a divergence is expected in $2+1$ dimensions of the form
\beq
\chi_{A} \; \mathrel{\mathop\sim_{g  \rightarrow g_c}} \; 
\vert g - g_c \vert^{- \alpha }
\label{eq:chi_sing}
\eeq
with exponent $\alpha \equiv 1 - \delta = 2 - 3 \nu $,
where $ \delta $ is the universal curvature exponent defined previously in
Eq.~(\ref{eq:delta_exp}),
and $\nu$ the correlation length exponent.
The latter is defined in the usual way \cite{par81,car97} through
\beq
\xi \; \mathrel{\mathop\sim_{g  \rightarrow g_c}} \;
\vert g - g_c \vert^{-\nu } \; ,
\label{eq:nu_exp}
\eeq
where $\xi$ is the invariant gravitational correlation length.
The scaling relations among various exponents ($\nu, \delta , \alpha$)
are rather immediate 
consequences of the scaling assumption for the singular part of the free
energy, $F_{sing} \sim \xi^{-d} $ in the vicinity of a critical point 
(for more detailed discussion see, for example, \cite{hbook,par81,car97}).
The preceding argument then implies, via scaling, that
a determination of $\alpha$ provides a direct estimate for
the correlation length exponent $\nu$ defined in Eq.~(\ref{eq:nu_exp}).
Note that based on the results so far one would be inclined
to conclude that for $2+1$ gravity the critical point
$g_c \rightarrow 0$ as $n \rightarrow \infty$.
Equation (\ref{eq:chi_sing}) can then be rewritten either as
\beq
\chi_{A} \; \mathrel{\mathop\sim_{g  \rightarrow g_c}} \; \xi^{\alpha / \nu }
\label{eq:chi_sing1}
\eeq
or, in a finite volume with linear lattice dimensions 
$L \sim N_0^{1/ d} \sim \sqrt{N_\Delta} \sim \sqrt{n} $ (since $N_\Delta=4n+2$), as
\beq
\chi_{A} \; \mathrel{\mathop\sim_{g  \rightarrow g_c}} \; 
L^{\alpha /  \nu } \; \sim \; n^{ 1 / \nu - 3/2 }  \; ,
\label{eq:chi_sing2}
\eeq
since, for a very large box and $g$ very close to 
the critical point $g_c$, the correlation length saturates
to its maximum value $\xi \sim L$.
Hence the volume- or $n$-dependence of $\chi$ provides
a clear and direct way to estimate the critical correlation length exponent
$\nu$ defined in Eq.~(\ref{eq:nu_exp}).

\vskip 40pt

\section{Results for Arbitrary Euler Characteristic $\chi$}
\label{sec:euler}

The results of the previous sections refer to regular triangulations
of the sphere ($\chi=2$) and the torus ($\chi=0$) in $2+1$ dimensions.
It would seem that one has enough information at this point to
reconstruct the same type of answers for arbitrary $\chi$.
In particular one has for the parameter $\beta$ 
[see Eqs.~(\ref{eq:beta_chi}) and (\ref{eq:beta_g_chi})]
\beq
\beta \; = \; { 2 \pi \chi \over g^3 } \; ,
\eeq
relevant for the wave functions in Eqs.~(\ref{eq:wavef_R}) or (\ref{eq:wavef_coul}).
For the average total area one then finds, using
the wave function expansion in Eq.~(\ref{eq:wavef_trunc}),
\beq
< A_{tot} >  =  g \, \left \{
 \frac{2^{1-2 n} \Gamma
   \left(n-\frac{1}{2}\right) \Gamma \left(2
   n+\frac{1}{2}\right)}{\Gamma (n)^3}
 + \frac{8 (n+1) \pi \, \chi \left [ 1-\frac{4^{1-2 n} \Gamma
   \left(n-\frac{1}{2}\right) \Gamma \left(n+\frac{1}{2}\right)
   \Gamma \left(2 n+\frac{1}{2}\right)^2}{n^2 \Gamma
   (n)^6}\right ] }{g^3 \, (2 n+1) } \, + \cdots
\right \} .
\label{eq:avar_chi}
\eeq
In the large $n$ limit one obtains for the average area of
a single triangle
\beq
< A_{\Delta} >  \; = \; 
{ g \over 2 \, \sqrt{ 2 \pi \, n } } \, 
\left [ 1 \, - \, { 5 \over 16 n } \, + \, 
{\cal O} \left ( { 1 \over n^2 } \right ) \right ]
\, + \, { (\pi-2 ) \, \chi \over g^2 \, n } \, 
\left [ 1 \, + \, { 1 \over 4 n ( \pi - 2) }
\, + \, \cdots  \right ] \; ,
\label{eq:avar_chi1}
\eeq
and for the average total area
\beq
< A_{tot} >  \; \sim \; 
\sqrt{ 2 \, n \over \pi } \, g \, + \, { 4 (\pi-2 ) \, \chi \over g^2 }
\, + \, \cdots  \; .
\label{eq:avar_chi2}
\eeq
For the area fluctuation defined in Eq.~(\ref{eq:chi_a1}) one finds
in the same large $n$ limit
\beq
\chi_A \; = \; \left ( 1 - { 2 \over \pi } \right ) \, { g^2 \over 4 }
\, + \, {\cal O} \left ( { 1 \over n } \right ) 
\, + \, ( 4 - \pi ) \sqrt{2 \over n \, \pi } \, { \chi \over g } 
\, + \, \cdots  \; .
\label{eq:chi_a2}
\eeq
Again note that the fluctuation appears to diverge as 
$g \rightarrow 0 $,
which implies that this is the more interesting limit, so
from now on we will focus specifically on this limit.
It is clear from the analytic expression for $ < A_{tot} > $ 
in Eqs.~(\ref{eq:avar_chi}) or (\ref{eq:avar_chi1}) 
that as $n \rightarrow \infty $, the gravitational
coupling $g(n)$, to this order in the Bessel expansion, has to scale like
\beq
g(n) \; \sim \; { 1 \over \sqrt{n} }  \; ,
\eeq
so that the expression for $ < A_{tot} > $ scales like $n$ or
$N_\Delta$, with the expression for $ < A_{\Delta} > $ staying finite.

The result of Eq.~(\ref{eq:chi_a2}) for $\chi_A $ then implies
\beq
\chi_A \; \sim \; { 1 \over g \, \sqrt{n} } \; \sim \; n^0
\eeq
in the same limit $ n \rightarrow \infty $.
In view of Eqs.~(\ref{eq:avar_chi}) and (\ref{eq:chi_sing2}) 
with $n \sim N_{\Delta} \sim L^2 $,
this would imply $ 2 / \nu - 3 = 0$, and thus for the
universal critical exponent $\nu$ itself 
$ \nu \; = \; { 2 \over 3 } \, = \, 0.666 $
to first order ($m=1$) in the Bessel function expansion of
Eq.~(\ref{eq:wavef_trunc}) and
$ \nu \; = \; { 17 \over 10 } \, = \, 0.588 $ 
to the next order ($m=2$) in the same expansion.


With some additional work one can, in fact, 
completely determine the asymptotic behavior of various averages
for large $\beta$ (small $g$) and large $n$.
First one notes that when $m$ Bessel functions are included 
in the expansion for
the wave function given in Eq.~(\ref{eq:wavef_trunc}),
beyond the leading order one at strong coupling, 
one obtains a wave function which contains powers of $\beta$ up to $\beta^m$.
For a given fixed $m$ one then finds for the average area per triangle the
following asymptotic result 
\beq
< A_\Delta >  \;\;  \sim \; { 1 \over g^{3 m - 1} \, n^{ m+1 \over 2} } \; ,
\label{eq:a_n}
\eeq
up to terms which contain higher powers of $1/n$ (making these less 
relevant in the limit $n \rightarrow \infty$), and also 
up to terms which are less singular in $g$ for small $g$.
The requirement that the average area per triangle be finite
as $n \rightarrow \infty$ then requires that the coupling $g$
itself should scale with $n$ according to
\beq
g(n) \; \sim \; { 1 \over n^{ m + 1 \over 2 ( 3 m -1) } }  \; .
\label{eq:g_n}
\eeq
For the area fluctuation itself one then computes in the same limit 
\beq
\chi_A \; \sim \; { 1 \over g^{3 m -2} \, n^{ m \over 2 } } \; ,
\label{eq:chi_gn}
\eeq
again to leading order in $1/n$ and $1/g$.
The requirement that $g(n)$ scale according to Eq.~(\ref{eq:g_n})
then implies from Eq.~(\ref{eq:chi_gn}) that the area fluctuation 
diverges in the limit $n \rightarrow \infty$ as
\beq
\chi_A (n) \; \sim \;  n^{ m - 1 \over 3 m - 1 } \; .
\label{eq:chi_n}
\eeq
By comparing with Eqs.~(\ref{eq:chi_sing1}) and (\ref{eq:chi_sing2})
one obtains immediately for the exponent
\beq
{ \alpha \over \nu } \, = \, { 2 \, m - 2 \over 3 \, m - 1 } \; ,
\eeq
and, therefore, from the scaling relation $\alpha = 2 - 3 \, \nu $
finally
\beq
\nu \, = \, { 6 \, m - 2 \over 11 \, m - 5 } \; .
\label{eq:nu_m}
\eeq
One can now take the limit $m \rightarrow \infty $ 
[infinite number of Bessel functions retained in the expansion of 
Eq.~(\ref{eq:wavef_trunc})], which leads to the exact result 
for the correlation length exponent of $2+1$ dimensional
quantum gravity
\beq
\nu \, = \, { 6 \over 11 } \, = \, 0.5454... \; .
\label{eq:nu_exact}
\eeq
The derivation shows that the exponent $\nu$ does not seem to depend on 
the Euler characteristic $\chi$ and, therefore, on the boundary
conditions.\footnote{One might wonder if the value for $\nu$ 
is affected by the choice of normalization in Eqs.~(\ref{eq:normaliz}) 
and (\ref{eq:measure}). It is easy to check that at least 
the inclusion of a weight factor $A^m$, with $m$ integer, 
does not change the result given in Eq.~(\ref{eq:nu_exact}).}
Furthermore one can compare the above value for $\nu$ with the 
(numerically exact) Euclidean three-dimensional quantum gravity result
obtained over twenty years ago in \cite{hw93}, namely $\nu \simeq
0.59(2)$.
It would, of course, be of great interest to repeat the above Euclidean
lattice calculation in order to refine the estimate 
and improve on the statistical and systematic uncertainty.
The exponent $\nu$ is expected to represent a universal quantity, independent
of short-distance regularization details and, therefore, characteristic of
gravity's universal scaling properties on 
distances much larger than the lattice cutoff.
As such, it should apply equally to both the Lorentzian and the
Euclidean formulation, and our results are consistent with this conclusion.
Moreover, in $3+1$ dimensions the exponent $\nu$ is a key physical
quantity as it determines the power for the running of
the gravitational
constant $G$ \cite{hat11} and for the Euclidean theory it is known \cite{ham00}
that the universal scaling exponent is consistent with $\nu = 1/3$.


It is perhaps worthwhile at this point to compare with other attempts
at determining the critical exponent $\nu$ in three-dimensional
gravity.
The latest and best results for quantum gravity in the perturbative
diagrammatic $2+\epsilon$ continuum expansion using the background
field method \cite{aid97,eps} give in $d=3$ ($\epsilon=1$ and
central charge $c=1$)
\beq
\nu^{-1} = 1 + {3 \over 5} + \dots
\eeq
to two-loop order and, therefore, $\nu \approx 0.625$, with
a substantial uncertainty of about fifty percent 
(which can be estimated for example by comparing the one- and two-loop results).
On the other hand, truncated renormalization group calculations for gravity
directly in three dimensions \cite{lit04,reu98} give to lowest order in the truncation
(i.e. with the inclusion of the cosmological and Einstein-Hilbert
terms only) the estimate
\beq
\nu^{-1} = { 2d (d-2)\over d+2 }
\eeq
and, therefore, in $d=3$ the value $\nu \approx 0.833$.
This last result is also affected by a rather substantial uncertainty
(again as much as fifty percent), which can be estimated, for example, 
by including curvature-squared terms in the truncated expansion.
Nevertheless, and in light of the uncertainties associated with the
various methods, it is very encouraging to note that widely different 
calculations (on the lattice and in the continuum) give
values for the universal scaling exponent $\nu$ that are 
roughly in the same ballpark.

From Eq.~(\ref{eq:nu_exact}) one obtains the fractal dimension for 
a gravitational path in $2+1$ dimensions
\beq
\nu^{-1} \, = \, d_F \, = \, { 11 \over 6 } \; = \; 1.8333...
\eeq
This is slightly smaller than the value for a free scalar field $d_F =2 $,
corresponding to the Brownian motion (or Wiener path) value.
It is closer to the value expected for a dilute branched polymer in the
same dimension \cite{zin98,cli10}, and the best match at this point
seems to be the $O(n)$ vector model for $n=-1$.
The exact value $\nu=6/11$ for $2+1$ gravity would then suggest a 
connection between the ground state properties of quantum
gravity and the geometry of dilute branched polymers in the same dimension.

In light of the results obtained so far it is possible to make a number
of additional observations.
First, note from Eq.~(\ref{eq:g_n}) that as $n \rightarrow \infty$,
the critical point (or renormalization group ultraviolet fixed point) 
moves to $g=0$
\beq
g(n) \; \mathrel{\mathop\sim_{ m \; \rightarrow \; \infty } } \; 
{ 1 \over n^{1/6} } \; .
\label{eq:gc1}
\eeq
For comparison, a variational calculation based on correlated product
(Slater-Jastrow) wave functions \cite{hw11} in $2+1$ dimensions gave
\beq
g_c^3 \; = \; 
\frac{ 4 \, \pi \; \chi }
{ N_\Delta \, \sqrt{ \sigma_0 ( \sigma_0 -2 ) } } \; ,
\label{eq:gc_var}
\eeq
where $\sigma_0 > 2 $ was a parameter associated there with the choice
of functional measure over edges.
The variational result of Eq.~(\ref{eq:gc_var}) can be compared
directly with the result of Eqs.~(\ref{eq:gc}) and (\ref{eq:gc1}),
for $\chi =2 $ and $N_\Delta= 2 \, n + 2 $.
Thus in both treatments the limiting value for the critical
point for $g$ in $2+1$ dimensions is zero, 
$g_c \rightarrow 0$ as the number of triangles $N_\Delta \rightarrow \infty$.

Physically, this last result implies that there is no weak coupling phase
($g < g_c $, or in terms of Newton's constant $G<G_c$):
the only surviving phase for gravity
in three dimensions is the strongly coupled one ($g > g_c$ or $G> G_c$).
Furthermore, the correlation length $\xi$ of Eq.~(\ref{eq:nu_exp}) is finite 
for $g>0$ and diverges at $g=0$.
In particular, the weak field expansion, which assumes $g$ small,
is expected to have zero radius of convergence.
\footnote{
These circumstances are perhaps unfamiliar in the gravity context, 
but are nevertheless rather similar to what happens in gauge theories,
including compact Quantum Electrodynamics in $2+1$ dimensions \cite{pol75}. 
There, the theory always resides in the strong coupling or
disordered phase, with a finite correlation length which eventually
diverges at zero charge.}
In a sense this is a welcome result, as in the Euclidean theory
the weak coupling phase was found to be pathological and thus physically unacceptable 
in both three \cite{hw93} and four dimensions \cite{hw84,ham00}.
It would seem, therefore, that the Euclidean and Lorentzian lattice results
are ultimately completely consistent: 
quantum gravity in $2+1$ dimensions always resides in
the strong coupling, gravitational antiscreening phase;
the weak coupling, gravitational screening phase is physically
excluded.
In addition, the exact value for $\nu$ determines, through standard 
renormalization group arguments, the scale dependence of the 
gravitational coupling in the vicinity of the ultraviolet fixed point \cite{hat11}.
\footnote{
Specifically, the universal exponent $\nu $ is related 
to the behavior of the Callan-Symanzik
beta function for Newton's constant $G$ in the vicinity of the ultraviolet fixed point by
$ \beta ' (G) \vert_{G=G_c} = - 1 / \nu $. 
Integration of the renormalization group equations for $G$ then
determines the scale dependence of $G (\mu) $ or $G( \Box )$ in
the vicinity of the ultraviolet fixed point.
Concretely, $\nu$ determines the exponent in the running of $G$.
One finds $ G ( \Box ) \, \sim \left ( \xi^2 \Box  \right )^{-1 / 2  \nu} $,
with $\Box \equiv g^{\mu\nu} \nabla_\mu \nabla_\nu$ the covariant
d'Alembertian and $\xi$ the renormalization group invariant correlation length.
A broader discussion of renormalization group methods as they
apply to quantum gravity can be found for example in \cite{hbook}.
}

\vskip 40pt

\section{Summary and Conclusions}
\label{sec:concl}

In this paper we have discussed the form of the gravitational
wave function that arises as a solution of the lattice Wheeler-DeWitt equation
[Eqs.~(\ref{eq:wd_latt}),(\ref{eq:wd_latt1}) and (\ref{eq:wd_2d})] for finite lattices.
The main result was the wave function $\Psi$ given in 
Eqs.~(\ref{eq:wavef_R}), (\ref{eq:wavef_coul}) and
(\ref{eq:wavef_R1}) with strong coupling limit (curvature term absent) corresponding to
the choice of parameter $\beta$=0.

To summarize, and for the purpose of the following discussion,
the wave function $\Psi$ given in Eq.~(\ref{eq:wavef_R1}) can be
written in the most general form as
\beq
\Psi \; \sim \; e^{- \, i\, x } \, 
_1F_1 \left( a, \, b, \, 2 \, i \, x \right)\; 
\label{eq:wavef_R2}
\eeq
up to an overall normalization constant $\tilde N$,
and with parameters related to various geometric invariants
\bea
a & \equiv &
\quarter \, N_{\Delta} \, - \, 
{ \sqrt{2} \, \pi \, i \over \sqrt{\lambda} \, G } \,\chi
\; = \;
\quarter \, N_{\Delta} \, - \, { i \over 2 \sqrt{ 2 \, \lambda} \, G } \,
\int d^2 y \, \sqrt{g} \; R
\nonumber \\
b & \equiv &
\half \, N_{\Delta}
\nonumber \\
x & \equiv &
{ \sqrt{ 2 \lambda } \over G } \, A_{tot} \; = \; 
{ \sqrt{ 2 \lambda } \over G } \, \int d^2 y \, \sqrt{g} \; .
\label{eq:wavef_args}
\eea
In the above definitions one can trade, if one so desires,  the total number of triangles
$N_\Delta$ for the total area
\beq
N_\Delta  \; = \; 
{ 1 \over < A_\Delta > } \, A_{tot} \; = \; 
{ 1 \over < A_\Delta > } \, \int d^2 y \; \sqrt{g} \; .
\label{eq:wavef_args1}
\eeq
Use has been made of the relationship between various coupling
constants ($g, G, \beta , {\tilde{\lambda}}, \lambda$) to reexpress
the wave function $\psi$ in slightly more general terms, as a function
of the original couplings $\lambda$ and $G$ appearing in the original form of
the Wheeler-DeWitt equation [see for example Eqs.~(\ref{eq:wd_c})
and (\ref{eq:wd_c1})].
We did show that an equivalent form for the wave function $\Psi$ can be
given  in terms of Coulomb wave functions [see Eq.~(\ref{eq:wavef_coul})], with argument
\beq
\beta \; \equiv \;
{ \sqrt{2} \, \pi \, \chi \over \sqrt{\lambda} \, G }
\; = \;
{ 1 \over 2 \sqrt{ 2 \, \lambda} \, G } \,
\int d^2 x \, \sqrt{g} \; R
\eeq
and $x$ defined as in Eq.~(\ref{eq:wavef_args}).

The above wave function is exact in the
limit of large areas and completely independent of the weak
field expansion.
Nevertheless it is only correct to some low order in the same
expansion in the limit of small areas.
This situation was interpreted as follows. 
For large areas one has a very large number of triangles, and the
short distance details of the lattice setup play a vanishingly small
role in this limit.
One recognizes this limit as being relevant for universal scaling properties,
including critical exponents.
For small areas on the other hand a certain sensitivity to the short
distance properties of the lattice regularization persists,
and thus a universal behavior is, not unexpectedly, hard
to achieve. 
In any case this last limit, in the absence of a truly fundamental
and explicit microscopic theory, is always expected to be affected by
short distance details of the regularization, no matter
what its ultimate nature might be (a lattice of some sort, 
dimensional regularization, or an invariant continuum momentum cutoff, etc.)


In principle any well-defined diffeomorphism-invariant average can be computed
using the above wave functions.
This will involve at some point the evaluation of a vacuum expectation
value of some operator $ {\cal \tilde O} (g) $
\beq
\langle \Psi \vert {\cal \tilde O} (g) \vert \Psi \rangle \; = \; 
{ \int  d \mu [g] \; {\cal \tilde O} ( g_{ij} ) \;
\vert \Psi [ g_{ij} ] \vert^2 
\over 
\int  d \mu [g] \; \vert \Psi [ g_{ij} ] \vert^2 }
\label{eq:exp_value}
\eeq
where $ d \mu [g] $ is the appropriate functional measure 
over three-metrics $g_{ij}$.
Evaluating such an average is, in general, non-trivial, as it requires
the computation of a (Euclidean) lattice path integral in one
dimension less 
\beq
\langle \Psi \vert {\cal \tilde O} (g) \vert \Psi \rangle \; = \; 
{\cal N} \int  d \mu [g] \; {\cal \tilde O} ( g_{ij} ) \;
\exp \left \{ - S_{eff} [g] \right \}
\label{eq:exp_value1}
\eeq
with $ S_{eff} [g] \equiv - \ln  \vert \Psi [ g_{ij} ] \vert^2 $ and
${\cal N}$ a normalization constant.
The operator $ {\cal \tilde O} (g) $ itself can be local, or nonlocal as
in the case of the gravitational Wilson loop discussed in \cite{hw07}.
Note that the statistical weights have many zeros corresponding to the nodes of the
wave function $\Psi$, and that $S_{eff}$ is infinite there.

In the previous sections we have shown that the wave function allows one to
calculate a number of useful and physically relevant averages and fluctuations,
which were later extrapolated to the infinite volume limit
of infinitely many triangles.
It was found that these diffeomorphism-invariant
observables point in $2+1$ dimensions to the existence of a fixed
point (a phase transition in statistical field theory
language) at the origin, $G_c =0 $.
One concludes, therefore, that the weak coupling (or gravitational
screening) phase has completely disappeared in the lattice nonperturbative
formulation and that the theory resides in the strong coupling phase only.
By contrast, in the Euclidean theory it was found in \cite{hw93} that
the weak coupling or gravitational screening phase exists but
is pathological, corresponding to a degenerate branched polymer.
A similar set of results is found in the four-dimensional Euclidean
theory, where the weak coupling, gravitational screening phase
also describes a branched polymer.
\footnote{
The nature of solutions to the lattice Wheeler-DeWitt equation
in $3+1$ dimensions will be discussed in a separate publication \cite{htw12}.}

The calculations presented in this paper and in \cite{hw11}
can be regarded, therefore, as consistent with the conclusions reached
earlier from the Euclidean framework, and
no new surprises arise when considering the Lorentzian $2+1$ theory.
Furthermore, we have emphasized before that the results obtained 
point at a nonanalyticity in the coupling
at $G=0$, signaling a strong vacuum instability of quantum gravitation
in this dimension.
In view of these results it is therefore not surprising
that calculations that rely on the weak field, semiclassical
or small $G$ expansion run into serious trouble and uncontrollable 
divergences very early on.
Such an expansion does not seem to exist if the non-perturbative lattice
results presented here are taken seriously.
The correct physical vacuum apparently cannot in any way be obtained
as a small perturbation of flat or near-flat spacetime.

Let us add here a few final comments aimed at placing the present work
in the context of previous calculations for the same theory.
A number of attempts have been made over the years to obtain an estimate
for the gravitational wave functional $\Psi [g]$ in the absence of sources.
These generally have relied on the weak field expansion in
the continuum, as originally done in \cite{kuc76,kuc92}.
Thus, for example, one finds in $3+1$ dimensions
\beq
\Psi [ h^{TT} ] \; = \; {\cal N} \; \exp \left \{ - \quarter \,
\int d^3 {\bf k} \; k \,
h^{TT}_{ik} ( {\bf k } ) \; h^{TT *}_{ik} ( {\bf k } ) \right \} \; ,
\eeq
where $h^{TT}_{ik} ( {\bf k } ) $ is the Fourier amplitude of
transverse-traceless modes for the linearized gravitational field
in four dimensions.
The above wave functional describes a collection
of harmonic oscillator wave functions, one for each of the infinitely
many physical modes of the linearized gravitational field.
As in the case of the electromagnetic field, the ground state
wave functional can be expressed equivalently
in terms of first derivatives of the field 
potentials (the corresponding ${\bf B}$ field for gravity),
without having to mention Fourier amplitudes, as
\beq
\Psi [ h^{TT} ] \; = \; {\cal N} \; \exp \left \{ - { 1 \over 8 \pi^2 } \,
\int d^3 {\bf x} \int d^3 {\bf y} \; 
{ h^{TT}_{ik,l} ( {\bf x } ) \; h^{TT *}_{ik,l} ( {\bf y } ) 
\over \vert {\bf x} \, - \, {\bf y} \vert^2 }
\right \} \; .
\eeq
Clearly both of the above expressions represent only
the leading term in an expansion involving infinitely
many terms in the metric fluctuation $h_{ij}$
(and since they apply to an expansion about flat space, the cosmological
constant term does not appear either).
Now, in $2+1$ dimensions the above expressions become meaningless,
since there cannot be any transverse-traceless modes.
The only expectation that remains true is that the wave functional
should still depend on physical degrees of freedom only:
it should be a function of the intrinsic geometry of 3-space
and should not change under a general coordinate change. 

If one restricts oneself to local terms a number of invariants
are possible in $2+1$ dimensions.
In principle, the wave function could depend on, 
besides the total area
\beq
A_{tot} \; = \; \int d^2 x \, \sqrt{g} 
\eeq
and curvature
\beq
4 \pi \, \chi  \; = \; \int d^2 x \, \sqrt{g} \; R  \; ,
\eeq
other invariants such as
\beq
r_n  \; = \; \int d^2 x \, \sqrt{g} \; R^n  \; 
\eeq
with $n$ an integer.
The latter result follows from the fact that in $d=2$, both the Riemann
and Ricci tensors only have one component, related to the scalar curvature,
\beq
R_{\mu\nu\rho\sigma} \; = \; \half \, R \, ( g_{\mu\sigma} \, g_{\nu\rho} 
\, - \, g_{\mu\rho} \, g_{\nu\sigma} ) \; ,
\;\;\;\;\;\;\;\;\;\;
R_{\mu\nu} \; = \; \half \; R \, g_{\mu\nu} \; .
\eeq
Nonlocal terms are possible as well involving inverse powers of the
covariant d'Alembertian $\Box$,
but these do not seem to play a significant role in the lattice theory.

Now, the relevant Euclidean theory for the present work is, of course,
gravity in three ($2+1$) dimensions.
But in three dimensions the Riemann and Ricci
tensor have the same number of algebraically
independent components (6) and are related to
each other by
\beq
R^{\mu\nu}_{\;\;\;\; \lambda \sigma} \; = \; 
\epsilon^{\mu\nu\kappa} \; \epsilon_{\lambda\sigma\rho} \; 
\left ( R^\rho_{\;\;\kappa} \, - \, \half \, \delta^\rho_{\;\;\kappa} \right )
\label{eq:riem-3d}
\eeq
The field equations then imply, using Eq.~(\ref{eq:riem-3d}),  
that the Riemann tensor is completely determined by the matter
distribution implicit in $T_{\mu\nu}$,
\beq
R_{\mu\nu\rho\sigma} = 8 \pi G \, \left [
g_{\mu\rho} \, T_{\nu\sigma}  +
g_{\nu\sigma} \, T_{\mu\rho}  +
g_{\mu\sigma} \, T_{\nu\rho}   -
g_{\nu\rho} \, T_{\mu\sigma}  +
T \, ( g_{\mu\sigma} \, g_{\nu\rho} 
- g_{\mu\rho} \, g_{\nu\sigma} ) \right ]
\eeq
In empty space $T_{\mu\nu}=0$, which then implies for zero
cosmological constant the vanishing of Riemann there
\beq
R_{\mu\nu\rho\sigma} \, = \, 0 \; .
\eeq
As a result in three dimensions classical spacetime is locally
flat everywhere outside a source, gravitational fields do not
propagate outside matter, 
and two bodies cannot experience any 
gravitational force: they move uniformly on straight lines.
There cannot be any gravitational waves either: the Weyl
tensor, which carries information about gravitational
fields not determined locally by matter, vanishes identically
in three dimensions.

What seems rather puzzling at first is that the Newtonian theory seems
to make perfect sense in $d=3$.
This can only mean that the Newtonian theory is {\it not} recovered
in the weak field limit of the relativistic theory.
To see this explicitly, it is sufficient to consider the trace-reversed form of the field equations,
\beq
R_{\mu\nu} \; = \; 8 \pi G \, \left ( \,
T_{\mu\nu} \, - \, { 1 \over d-2 } \; g_{\mu\nu} \, T  \,
\right )
\eeq
with $T = T^\lambda_{\;\; \lambda} $, in the weak field limit.
In the linearized
theory, with $h_{\mu\nu} = g_{\mu\nu} - \eta_{\mu\nu}$,
and in the gauge 
$ \nabla_\lambda h^{\lambda}_{\;\; \mu} - \half \nabla_\mu h^{\lambda}_{\;\;\lambda} =0 $,
one obtains the wave equation
\beq
\Box \, h_{\mu\nu} \; = \; - 16 \pi G \, \left ( \,
\tau_{\mu\nu} \, - \, { 1 \over d-2 } \; \eta_{\mu\nu} \, \tau  \,
\right )
\eeq
with $\tau_{\mu\nu}$ the linearized stress tensor.
After neglecting the spatial components of $\tau_{\mu\nu}$
in comparison to the mass density $\tau_{00}$,
and assuming that the fields are quasistatic, one
obtains a Poisson equation for $h_{00}$,
\beq
\nabla^2 \, h_{00} \; = \; - 16 \pi G \; { d-3 \over d-2 } \; \tau_{00}
\label{eq:box-h}
\eeq
In four dimensions this is equivalent to Poisson's equation for the Newtonian theory
when one identifies the metric with the Newtonian field $\phi$ in
the usual way via $ h_{00} = - 2 \phi $.
But in three dimensions such a correspondence is obstructed
by the fact that, from Eq.~(\ref{eq:box-h}), the nonrelativistic
Newtonian coupling appearing in Poisson's equation is given by
\beq
G_{\rm Newton} \; = \; { 2 \, (d - 3) \over ( d - 2 ) } \; G
\eeq
and the mass density $\tau_{00}$ completely decouples from
the gravitational field $h_{00}$.
As a result, the linearized theory in three dimensions
fails to reproduce the Newtonian theory.

In a complementary way one can show that
gravitational waves are not possible either in the linearized theory
in three dimensions.
Indeed the counting of physical degrees of freedom for the
$d$-dimensional theory goes as follows.
The metric $g_{\mu\nu}$ has $\half d(d+1)$ independent components;
the Bianchi identity and the coordinate conditions
reduce this number to $\half d(d+1) - d - d = \half d (d-3)$, which
gives indeed the correct number of physical degrees of freedom
(two) corresponding to a massless spin two particle in $d=4$,
and no physical degrees of freedom in $d=3$
(and minus one degree of freedom in $d=2$, which is in fact incorrect).
Nevertheless, investigations of quantum two-dimensional gravity have uncovered
the fact that there can be surviving degrees of freedom
in the quantum theory, at least in two dimensions.
The usual treatment of two-dimensional gravity \cite{pol81}
starts from the metric in the conformal gauge
$g_{\mu\nu} (x) = e^\phi (x) \tilde{g}_{\mu\nu}$, where 
$\tilde{g}_{\mu\nu}$ is a reference metric, usually
taken to be the flat one.
The conformal gauge-fixing then implies a nontrivial
Faddeev-Popov determinant, which, when exponentiated, results
in an effective Liouville action, with a potential term coming from 
the cosmological constant contribution. 
One would, therefore, conclude from this example that gravity
in the functional integral representation needs a careful
treatment of the conformal degree of freedom, since
in general its dynamics cannot be assumed to be trivial.
The calculations presented in this paper show that this
is indeed the case in $2+1$ dimensions as well.


\vspace{20pt}

{\bf Acknowledgements}

The work of H.W.H was supported in part by the University of California.
The work of R.M.W. was supported in part by the UK Science and Technology Facilities Council. 
The work of R.T. was supported in part by a DED GAANN Student Fellowship.

\vskip 40pt


\vfill


\end{document}